\documentclass[letterpaper,twocolumn,10pt]{article}
\usepackage{usenix-2020-09}

\usepackage{amsmath}

\usepackage{graphicx}

\usepackage{booktabs}

\usepackage{placeins}

\usepackage{xcolor}

\usepackage{enumitem}
\setlist[itemize,1]{
  leftmargin=\parindent,
  labelsep=0.45em,
  itemsep=2pt,
  topsep=4pt,
  parsep=0pt,
  partopsep=0pt
}

\usepackage{algorithm}

\usepackage[
  noEnd=false,
  indLines=true,
  rightComments=true,
  italicComments=false,
  commentColor=blue
]{algpseudocodex}

\makeatletter
\newcommand{\algphase}[1]{%
  \algpx@endCodeCommand%
  \Statex
  \textcolor{blue}{\textbf{#1}}%
}
\makeatother

\usepackage{pifont}

\usepackage{xurl}
\newcommand{\heading}[1]{%
  \par
  \addvspace{.4em}%
  \noindent\textbf{#1}\nobreak\hspace{0.5em}\ignorespaces
}

\newlist{requirements}{enumerate}{1}
\setlist[requirements,1]{
  label=\textbf{R\arabic*:},
  ref=R\arabic*,
  wide=0pt,
  labelsep=0.5em,
  topsep=0.25\baselineskip,
  itemsep=0.25\baselineskip,
  parsep=0pt,
  partopsep=0pt
}

\newcommand{\sysname}{CadenceRL}

\newcommand{\affmark}[1]{%
  \textsuperscript{\ensuremath{#1}}%
}

\newcommand{\affone}{\affmark{1}}
\newcommand{\afftwo}{\affmark{2}}

\begin{document}

\date{}

\title{\Large \bf
  Reshaping Rollout Workloads for Asynchronous RL Post-Training\\
  on Heterogeneous Accelerators
}

\author{%
  {\rm Jiahui Li}\affone,
  {\rm Hao Nie}\afftwo,
  {\rm Yibo Zhu}\afftwo,
  {\rm Pengjin Xie}\affone\\
  {\rm Yu Zhou}\afftwo,
  {\rm Xiaolong Zheng}\affone,
  {\rm Liang Liu}\affone,
  {\rm Huadong Ma}\affone\\[0.15em]
  \affone\hspace{0.15em}Beijing University of Posts and Telecommunications
  \hspace{1.5em}
  \afftwo\hspace{0.15em}StepFun%
}

\maketitle

\begin{abstract}

Reinforcement learning (RL) post-training
increasingly relies on long-horizon, multi-turn rollouts.
As post-training jobs outgrow a single cluster,
rollout pools assembled across clusters
introduce hardware heterogeneity.
Rollout scheduling must serve two stakeholders:
the hardware needs high aggregate decode throughput,
while each trajectory needs to finish quickly.
The tension arises
from the memory-bandwidth-bound nature
of autoregressive decoding.
A large active batch
amortizes weight reads for high throughput
but leaves each trajectory
a smaller bandwidth share
and a longer completion time.
The scheduling objective
is therefore specialization,
letting different workers
serve different roles.
Heterogeneous hardware further enables this specialization.
High-bandwidth accelerators
favor long-context work, 
while cost-efficient accelerators 
sustain large batches.
Workload evolution
makes this specialization difficult to sustain,
and dynamic reassignment
faces a circular dependency
because a move's benefit depends on 
subsequent placement decisions.

We present \sysname{},
which bypasses this dependency through
structural workload reshaping
rather than per-move benefit estimation.
Pacing replaces long-context trajectories with shorter ones, 
providing a structurally positive transformation that 
sustains large active batches
for high throughput.
When accumulated staleness demands faster completion,
concentration directs the residual long-context tail 
onto high-affinity workers.
Late-bound KV preparation 
stages accumulated prefixes 
before a destination is selected.
On heterogeneous rollout pools,
\sysname{} improves decode throughput
by up to 48\%
and reduces P95 trajectory latency
by up to 64\%.
Adding high-bandwidth accelerators
reduces tail latency,
while adding cost-efficient accelerators
increases throughput,
without manual routing configuration.

\end{abstract}


\section{Introduction}
\label{sec:introduction}

With the emergence of long-horizon scenarios,
large language models (LLMs) are expected to reason
over extended contexts and interact
with external environments over multiple turns,
rather than only generate a single response~\cite{yao2022react,wang2024survey}.
Reinforcement learning (RL) post-training
has become an important paradigm
for scaling the reasoning capabilities of LLMs
across challenging domains,
including mathematics~\cite{guo2025deepseek},
coding~\cite{jimenez2024swe},
and other tasks~\cite{rein2023gpqa,hendrycks2020measuring,yue2024mmmu}.
As models and tasks grow,
the compute demand of a post-training job
can exceed the usable accelerator capacity
of a single cluster.
Large post-training jobs therefore increasingly need
to combine accelerator capacity
across multiple clusters.

Recent asynchronous RL post-training systems
address this capacity challenge
by disaggregating rollout (trajectory collection)
from training.
Rollout is dominated by autoregressive decoding
and memory bandwidth,
whereas training requires high compute throughput
for policy updates.
The two stages run concurrently
in independently provisioned resource pools,
exchanging trajectories and policy weights
between pools~\cite{wu2026weave,wang2026dynarl,he2026hetrl,tan2026dynamic}.

In practice,
capacity assembled from multiple clusters
is rarely homogeneous.
Accelerator generations overlap
during incremental upgrades,
procurement constraints lead operators
to combine devices from different vendors,
and cost considerations
favor mixed deployments~\cite{meta_nvidia_2026,anthropic_amazon_2026}.
Older devices remain productive
while newer capacity is added,
making this heterogeneity persistent.
Disaggregation resolves the mismatch
between rollout and training
by placing each stage
in a dedicated resource pool.
However, the rollout pool itself
may span accelerator types
with different compute capability,
memory bandwidth, and cost per device-hour,
and trajectory placement
within this heterogeneous pool
remains unresolved.

Efficient rollout scheduling must serve
two stakeholders with different needs.
The hardware needs high aggregate decode throughput, 
which is the pool's total token generation rate.
Each trajectory needs low completion time, 
especially the long-context trajectories
that determine when the policy version can advance,
since a trajectory that lingers
grows stale against the updating policy.
These two needs draw on the same physical resource, memory bandwidth, 
but prefer opposite ways of using it.
The tension arises
from the memory-bandwidth-bound nature
of autoregressive decoding.
Each decode step reads model weights
shared across the active batch
and accesses per-trajectory key-value (KV) cache state,
both through the same memory bandwidth.
A large active batch amortizes
the fixed cost of weight reads
across many trajectories,
giving the hardware high aggregate decode throughput.
But a large batch also means
each trajectory receives a smaller share
of the worker's memory bandwidth
and decodes more slowly.
A worker cannot serve both needs at once:
the batch must be large for throughput
or small for fast completion.
The scheduling problem is therefore
not to compromise on every worker,
but to let different workers
serve different objectives---%
a division we call \emph{specialization}.

\begin{figure}[t]
  \centering
  \includegraphics[width=\columnwidth]{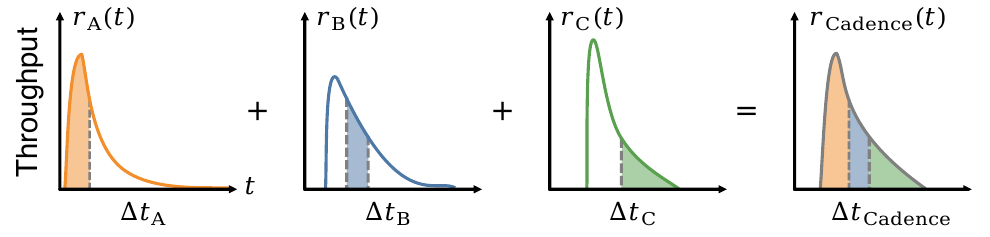}
  \caption{%
  \textbf{Complementary specialization
  across heterogeneous accelerators.}
  The four plots show the throughput profiles
  $r_\mathrm{A}(t)$, $r_\mathrm{B}(t)$,
  $r_\mathrm{C}(t)$, and $r_{\mathrm{Cadence}}(t)$
  over their respective rollout lifecycles.
  Accel-A and Accel-B contribute high throughput
  in the large-batch, short-context phase
  but decline sharply as contexts grow;
  Accel-C maintains a widening advantage
  in the long-context region.
  \sysname{} coordinates these complementary strengths
  to achieve higher aggregate throughput
  and a shorter drain time than any fixed assignment.}
  \label{fig:introduction:objective}
\end{figure}

Heterogeneous hardware presents
an opportunity to deepen this specialization.
High-bandwidth accelerators
are naturally suited
to concentrate decode resources
on a few long-context trajectories,
while cheaper, lower-bandwidth accelerators
can cost-effectively sustain
large active batches of shorter contexts.
Hardware diversity is therefore not merely
a source of mismatch to be managed,
but the natural foundation
for complementary specialization.
Figure~\ref{fig:introduction:objective}
illustrates this complementarity.

The difficulty is sustaining this specialization
as the workload evolves.
Rollout trajectories vary in response length,
turn count, and final context length,
all unknown at dispatch
and determined by generation
and environment interactions.
As shorter trajectories complete,
the active batch shrinks
while surviving contexts grow,
shifting each worker away
from its initial operating point.
A worker that initially sustained
a large active batch at high throughput
may end up serving only a few long-context residents. 
It loses its throughput advantage
without gaining the concentrated bandwidth
that benefits those remaining trajectories.
Specialization must therefore be
continuously maintained,
not assigned once at dispatch.

Sustaining specialization
requires dynamic trajectory reassignment
that moves work between workers
as the workload evolves.
A natural approach is to guide each reassignment
by its predicted benefit,
but a move's benefit depends
on the trajectory's unknown remaining length
and the destination's future batch composition,
which is itself shaped
by subsequent placement decisions, 
creating a \emph{circular dependency}
between estimating benefits
and choosing placements.
Existing systems sidestep this difficulty
through simplifying assumptions:
pool-level systems assume
homogeneous rollout workers~\cite{wu2026weave};
resource-allocation systems adjust capacity
across pipeline stages
without reorganizing trajectories
within a rollout pool~\cite{wang2026dynarl};
hardware-aware planners rely on
static device specifications
rather than dynamic
workload-driven redeployment~\cite{he2026hetrl}.
These simplifications avoid the circular dependency
but forgo the specialization opportunity
within the heterogeneous pool.

We present \sysname{},
a system for heterogeneous rollout scheduling
in RL post-training.
The decode-throughput landscape
offers a structural way past this difficulty.
Throughput improves monotonically
as long-context work departs
and shorter work arrives,
so reshaping toward specialization
has guaranteed positive gain.
This guarantee lets the scheduler act
without per-migration benefit estimation,
bypassing the circular dependency entirely.
Under \sysname{},
the pool differentiates accordingly:
\emph{throughput workers}
sustain large active batches
in the high-throughput region,
while \emph{tail workers}
dedicate concentrated decode resources
to long-context trajectories.

\sysname{} manages the throughput--latency trade-off
by dynamically adjusting
the ratio between these two roles
through two mechanisms
that share a common control structure
in which departure and placement are separated in time
and connected through KV headroom.
While fresh work remains available,
the \emph{long-out, short-in} transformation
keeps workers in higher-throughput states for longer.
Specialization emerges from this process:
workers that absorb short replacements
become throughput workers,
while workers that accumulate long-context work
become tail workers.
When the staleness constraint
exhausts the supply of fresh work,
the scheduler concentrates
the residual tail onto high-affinity workers,
freeing weaker workers
for fresh work from the next policy version
and restoring the conditions for pacing.
The staleness budget governs
the transition between these phases
and thereby the throughput--latency balance.

We also design a KV preparation mechanism
that supports deferred destination binding.
When a trajectory departs,
its accumulated KV prefix is staged
before a destination is selected.
A compatible destination within the same cluster
retrieves the prepared state
without recomputing the prefix;
other migrations reconstruct it from tokens.
Preparation proceeds independently of placement,
allowing the scheduler to choose destinations
from the latest workload state.

This paper makes the following contributions:
\begin{itemize}
    \item We reframe the throughput--latency relationship
    in autoregressive decoding
    and show that high throughput
    and fast tail completion
    require different workload compositions
    rather than a compromise on every worker,
    identifying specialization as the scheduling objective.

    \item We show that the decode-throughput landscape's
    monotonicity guarantees positive gain
    from reshaping toward specialization,
    letting the scheduler act
    without pairwise benefit estimation.
    
    \item We design two scheduling mechanisms
    that share a control structure based on KV headroom.
    Pacing sustains throughput
    through tight-fit eviction and replacement
    from which specialization emerges
    without explicit role assignment.
    Concentration consolidates the residual tail
    onto high-affinity workers
    when fresh work is exhausted.
    
    \item We design late-bound KV preparation
    that stages accumulated prefixes
    before a destination is selected,
    preserving deferred binding
    without prefix recomputation at the target.
    
    \item We implement \sysname{}
    and evaluate it
    on heterogeneous rollout pools.
    On heterogeneous configurations,
    \sysname{} improves aggregate decode throughput
    by up to 48\%
    while reducing P95 trajectory latency
    by up to 64\%
    compared to the strongest baseline.
    Heterogeneous scaling is composable. 
    Adding high-bandwidth accelerators
    reduces tail latency,
    while adding cost-efficient accelerators
    increases throughput,  
    without manual routing configuration.
\end{itemize}

\section{Background}
\label{sec:background}

\subsection{RL Post-Training Workflow}
\label{sec:background:workflow}

\begin{figure}[t]
  \centering
  \includegraphics[width=\columnwidth]{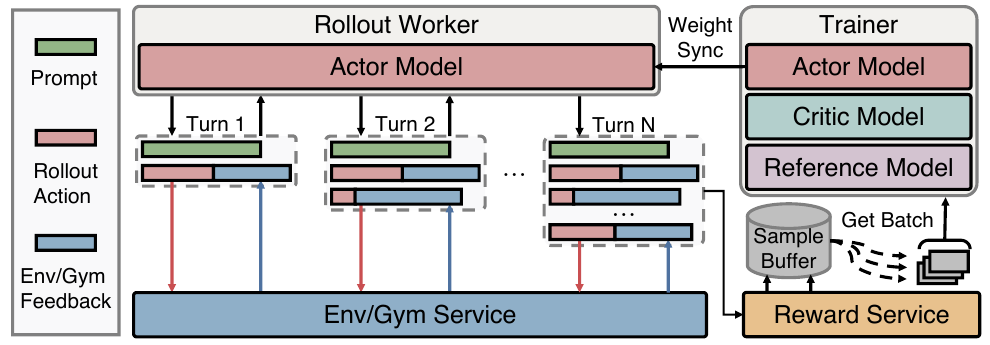}
  \caption{RL workflow in a disaggregated architecture.}
  \label{fig:background:workflow}
\end{figure}

\heading{The training loop.}
RL post-training alternates between
trajectory collection and policy updates,
as shown in Figure~\ref{fig:background:workflow}.
During rollout,
the model interacts with an environment
over one or more turns
to produce a trajectory.
Completed trajectories are scored
by task-specific reward signals
and consumed by the trainer,
which updates the policy
and distributes the new weights
to the rollout workers.

\heading{Asynchronous execution.}
Rollout is dominated by autoregressive decoding
and primarily constrained by memory bandwidth,
whereas training requires high compute throughput
for weight updates.
Asynchronous systems place the two stages
in separately provisioned resource pools,
so that rollout proceeds concurrently
with training~\cite{fu2026areal}.

\heading{Policy versioning.}
Policy-gradient algorithms
such as PPO~\cite{schulman2017proximal}
and GRPO~\cite{shao2024deepseekmath}
require each trajectory to be generated
under a fixed behavior policy.
Each worker therefore serves
one policy version at a time
and can adopt newer weights only after
all resident trajectories
have completed or departed.
A worker that drains quickly
resumes fresh work under the new version;
one held back by long-running trajectories
cannot advance
and remains in a low-throughput state.
These lingering trajectories
accumulate policy staleness---%
the gap between their collection version
and the current training version.
A long-running tail
is therefore both a throughput bottleneck
and a source of stale training data.

\subsection{Workload Evolution}
\label{sec:background:workload}

\begin{figure}[t]
  \centering
  \includegraphics[width=\columnwidth]{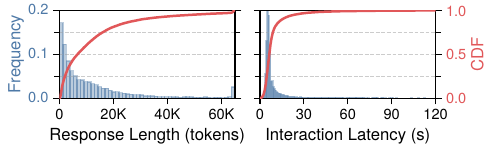}
  \caption{%
  Trajectory length and interaction latency distributions in Qwen3-8B post-training traces.}
  \label{fig:background:workload-distribution}
\end{figure}

\begin{figure*}[t]
  \centering

  \begin{minipage}[t]{0.245\textwidth}
    \setlength{\abovecaptionskip}{-12.5pt}%
    \vspace{2pt}%
    \centering
    \includegraphics[width=\linewidth]{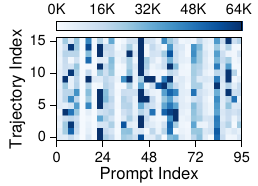}
    \caption{Within-prompt variation in trajectory-level response length.}
    \label{fig:background:response-variation}
  \end{minipage}%
  \hspace{0.02\textwidth}%
  \begin{minipage}[t]{0.735\textwidth}
    \vspace{0pt}%
    \centering
    \includegraphics[width=\linewidth]{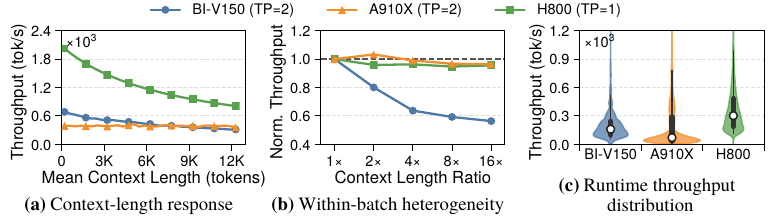}
    \caption{Workload-dependent Qwen3-8B decode throughput across configured accelerator backends, shown as controlled profiles and runtime distributions.}
    \label{fig:background:hardware-response}
  \end{minipage}
\end{figure*}

The workload served by each rollout worker
evolves throughout an iteration
as trajectories complete and leave the active batch
while surviving trajectories accumulate context.
The pace of this evolution
depends on the individual trajectories
the worker hosts,
and a trajectory's demand---%
response length, turn count,
and wall-clock duration---%
is determined at runtime.
In a coding task, for instance,
the agent generates a solution at each turn
and submits it to a sandboxed environment
for compilation and testing.
A successful submission terminates the trajectory.
Failure triggers revision,
extending the context with environment feedback.
Whether the trajectory continues,
how long each response is,
and how long the environment takes to respond
are all decided
by the model--environment interaction at each step.
Figure~\ref{fig:background:response-variation}
shows that trajectories from the same prompt
under the same policy
differ in response length
by an order of magnitude.
Across many prompts,
these differences aggregate
into the long-tailed distributions
of response length and interaction latency
shown in Figure~\ref{fig:background:workload-distribution}.
An initially balanced assignment
does not guarantee balanced progress,
as workers serving different trajectory mixes
diverge when some trajectories terminate early
while others enter extended interaction sequences.

\subsection{Hardware-Workload Interaction}
\label{sec:background:hardware}

Decode throughput depends on the combination
of active batch size and context lengths,
and this dependence varies across hardware.
The BI-V150~\cite{iluvatar_biv150},
A910X~\cite{huawei_a910x},
and H800~\cite{nvidia_h800} accelerators
in Table~\ref{tab:background:accelerator-specs}
differ in compute capability,
memory bandwidth,
and cost per device-hour.
Figure~\ref{fig:background:hardware-response}
shows how these differences
translate into workload-dependent throughput.
At a fixed active batch size,
Figure~\hyperref[fig:background:hardware-response]{%
  \ref*{fig:background:hardware-response}(a)}
shows that H800 sustains
the highest throughput across context lengths.
BI-V150 starts above A910X
but declines as context grows,
while A910X remains comparatively flat,
bringing the two closer at long contexts.
Figure~\hyperref[fig:background:hardware-response]{%
  \ref*{fig:background:hardware-response}(b)}
shows that this within-batch heterogeneity
affects hardware unevenly---%
a worker's effective throughput
depends not only on the mean context length
but on the composition of its resident batch.
Figure~\hyperref[fig:background:hardware-response]{%
  \ref*{fig:background:hardware-response}(c)}
shows that under one-time placement,
workers frequently operate
far from their controlled-profile throughput,
because the workload assigned at dispatch
has evolved since.
This growing mismatch
motivates dynamic workload redistribution.

\begin{table}[t]
  \centering
  \setlength{\abovecaptionskip}{4pt}
  \setlength{\belowcaptionskip}{5pt}
  \caption{Per-device hardware specifications and hourly costs.}
  \label{tab:background:accelerator-specs}
  \small
  \setlength{\tabcolsep}{4pt}
  \begin{tabular*}{\columnwidth}{l@{\extracolsep{\fill}}ccc}
    \toprule
      & \textbf{BI-V150} & \textbf{A910X} & \textbf{H800} \\
    \midrule
    Device type                     & GPU & NPU & GPU \\
    Devices per machine             & 16 & 16 & 8 \\
    BF16 compute (TFLOP/s)          & 96 & 313 & 989 \\
    HBM capacity (GB)               & 32 & 64 & 80 \\
    HBM bandwidth (GB/s)            & 800 & 1{,}935 & 3{,}350 \\
    Interconnect bandwidth (GB/s)   & 64 & 392 & 400 \\
    Cost (\$/h)                     & 0.22 & 0.67 & 2.00 \\
    \bottomrule
  \end{tabular*}
\end{table}

\section{Design Principles}
\label{sec:principles}

The workload served by a heterogeneous rollout pool
changes throughout an iteration,
with the active batch shrinking as trajectories complete
while surviving contexts grow.
These changes shift the relative advantage
of different accelerators,
creating a continuous need
to match workload to hardware.
\sysname{} addresses this through \emph{specialization}:
rather than optimizing every worker
for the same aggregate metric (throughput or latency),
the scheduler lets different workers
serve different objectives
according to their hardware characteristics
and current workload composition.

\subsection{Specialization in Heterogeneous Pools}
\label{sec:principles:specialization}

\heading{When low throughput is not waste.}
Rollout scheduling is often framed as a conflict
between throughput and latency.
Higher utilization increases aggregate throughput
at the expense of individual trajectory latency.
Under this view,
any idle capacity is waste to be eliminated---%
a principle taken to its extreme by partial-rollout strategies
that interrupt ongoing trajectories
to keep every worker fully occupied~%
\cite{fu2026areal,zhou2025april,qu2025copris}.

The decode-throughput landscape of autoregressive decoding
suggests a different perspective.
Let $B$ denote the active batch size
and $\bar{C}$ the mean context length of the active trajectories.
Decode throughput is highest in the large-$B$, short-$\bar{C}$ region,
where model-weight reads are amortized across many trajectories.
As $B$ shrinks and $\bar{C}$ grows,
KV-cache traffic approaches the memory-bandwidth limit
and decode throughput enters a plateau~%
\cite{pope2023efficiently,recasens2026slim}.

With fewer co-resident trajectories, however, 
each one claims a larger share of the worker's memory bandwidth
and decodes faster.
Low throughput for the worker
means low latency for the trajectory.
High aggregate throughput
and fast tail completion
therefore do not conflict---%
they require different workload compositions
that cannot coexist on the same worker.

\heading{Throughput workers and tail workers.}
The rollout pool can sustain both compositions simultaneously
by specializing workers into complementary roles.
\emph{Throughput workers} sustain large active batches
in the high-throughput region for as long as possible,
maximizing the time integral of decode throughput.
\emph{Tail workers} dedicate their decode resources
to long-context trajectories,
hosting as few co-residents as necessary
so that each trajectory completes quickly.
The two roles reinforce each other:
the longer throughput workers sustain their favorable composition,
the fewer trajectories remain for tail workers to host,
and the more concentrated their decode resources become.

\heading{Hardware complementarity.}
Heterogeneous hardware amplifies this opportunity.
Under long-context decode,
attention often becomes memory-bandwidth-bound,
making memory capacity and bandwidth
important workload-affinity dimensions~%
\cite{pope2023efficiently,recasens2026slim}.
These differences align naturally
with the workload roles above---%
high-bandwidth accelerators absorb long-context work,
while cost-efficient accelerators 
sustain large batches for throughput.
In a homogeneous pool the same specialization logic applies,
driven by workload composition rather than hardware,
but without hardware complementarity
to amplify the benefit.

\subsection{Circular Dependency in Benefit Estimation}
\label{sec:principles:circular}

A direct approach to realizing the specialization
would estimate the benefit
of moving each unfinished trajectory to each candidate worker
and migrate whenever the estimated gain is positive.
This estimation has a \emph{circular dependency}.
The gain of a migration depends on the trajectory's remaining length,
which is unknown and often heavy-tailed.
It depends on the other trajectories
that will later share the destination,
but those co-residents are determined by scheduling decisions not yet made.
Transfer and KV-prefix recomputation add costs
that depend on when and where the migration occurs.
The workload state used to evaluate one migration
therefore depends on migrations not yet chosen.
Scheduling decisions determine the future workload state,
yet that future state
is required to evaluate those decisions.

Pairwise benefit estimation is intractable
not because the computation is expensive,
but because the inputs to the computation
are logically unavailable.
If instead there exists a class of workload transformations
whose gain is guaranteed to be positive
by the structure of the problem,
regardless of specific remaining lengths,
future co-residents, or migration timing,
then the scheduler can act on this structural guarantee
without entering the estimation loop.

\subsection{Guaranteed Gain from Workload Reshaping}
\label{sec:principles:gain}

The decode-throughput landscape described above
provides such a class of transformations.
A composite transformation that keeps the workload
in higher-throughput states for longer
guarantees positive gain
by decoding more tokens over the same wall-clock time.

\begin{figure}[t]
  \centering
  \includegraphics[width=\columnwidth]{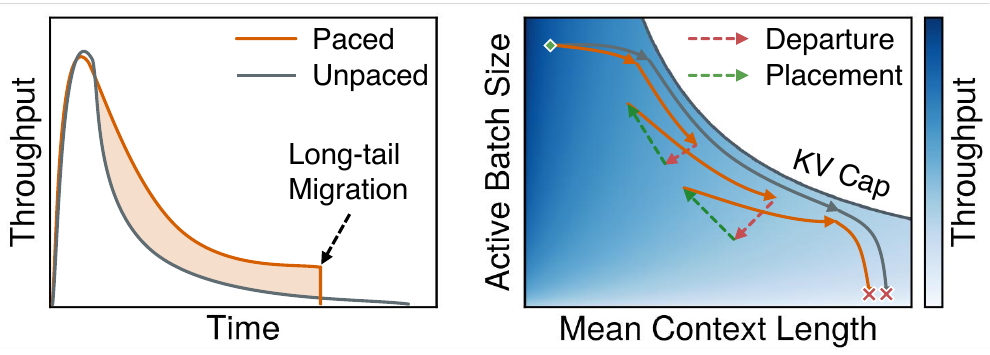}
  \caption{%
  Pacing the decode-throughput arc through workload reshaping.
  \textbf{Left:} throughput over time.
  \textbf{Right:} workload evolution
  in active batch size--mean context length space.
  Darker colors indicate higher throughput,
  and the slanted boundary marks the KV-capacity limit.
  Long-context departures and short-context placements
  jointly keep the workload
  in higher-throughput states for longer.}
  \label{fig:principles:throughput-arc}
\end{figure}

\heading{Pacing the throughput arc.}
While shorter-context work remains available,
a long-context resident can be removed from a worker
and replaced by one or more shorter-context trajectories
whenever the replacement fits within the available capacity.
This combined transformation---%
\emph{long out, short in}---%
simultaneously slows both components of the throughput decline:
the active batch remains larger for longer,
and the mean context length grows more slowly.
Figure~\ref{fig:principles:throughput-arc} shows that
the workload trajectory stays
in higher-throughput regions of the landscape for longer,
and the throughput arc declines more slowly
than the unpaced curve.
We call this effect \emph{pacing the throughput arc}.

The gain is guaranteed
because the transformation keeps the workload
in higher-throughput states,
not because any individual departure or placement
is independently beneficial.
Departure alone would reduce $B$,
accelerating the batch-size decline it aims to prevent.
Placement alone,
without removing a long-context resident,
would leave the existing context pressure in place.
Only their combination---%
long out followed by short in---%
simultaneously maintains $B$ and slows the growth of $\bar{C}$.

\heading{From pacing to specialization.}
The \emph{long-out, short-in} transformation
is directional because
replacing long-context residents with shorter ones
always moves the workload toward higher-throughput states.
At the same time,
the displaced long-context trajectories
leave their current workers but remain in the system,
where they must be served by other workers.
A directional transformation
applied to a fixed workload
necessarily produces differentiation:
workers that perform the replacement
become throughput workers;
workers that absorb the displaced trajectories
become tail workers.
The scheduler does not assign these roles explicitly.
They emerge from the transformation itself---%
a global reshaping of the pool's workload distribution
rather than a sequence of pairwise migration decisions.
When the replacement source is exhausted,
the composite can no longer form.
\S\ref{sec:principles:tradeoff} shows
that the guaranteed gain then takes a different form.

\subsection{The Throughput--Latency Trade-off}
\label{sec:principles:tradeoff}

\heading{When less active time means faster completion.}
A long-context trajectory displaced during pacing
alternates between active decode
and a pending state while waiting for capacity.
Its trajectory latency therefore includes
both decode time and time spent pending.
Without reshaping,
it may decode continuously,
but its worker's shrinking batch
and growing contexts dilute its bandwidth share,
so each token takes longer.
Pacing instead lets the trajectory resume
in concentrated windows with fewer co-residents.
These windows make faster progress
and can outweigh the time spent waiting,
so a trajectory may spend less time decoding
yet finish sooner.
Pacing controls both terms.
Sustaining favorable workload composition
raises pool throughput and speeds each resume window,
but can also lengthen the wait before that window.
 
\heading{Staleness as the scheduling knob.}
RL training introduces a natural resolution to this tension.
The staleness constraint bounds
the number of policy versions
that may overlap during training.
This bound limits
how much fresh short-context work
enters the system from newer policy versions,
and therefore how long pacing can sustain the short end.
A generous staleness budget extends the pacing phase,
increasing throughput
but prolonging displacement for the tail.
A strict budget exhausts the short end sooner,
reducing throughput
but accelerating tail completion.
 
\heading{From pacing to concentration.}
When the short end is exhausted,
the guaranteed gain does not disappear---%
it changes form.
The \emph{long-out, short-in} transformation
loses its replacement source
and can no longer sustain throughput through reshaping.
The scheduler instead concentrates
the residual long-context tail
onto workers best suited for long-context decode,
freeing the remaining workers
for new short-context work
that restores the conditions for pacing.
The guaranteed gain therefore persists across phases---%
in the pacing phase
it comes from sustaining throughput;
in the concentration phase
from accelerating tail drain.
Neither phase requires per-migration benefit estimation.

\section{\sysname{} Design}
\label{sec:design}

\begin{figure}[t]
  \centering
  \includegraphics[width=\columnwidth]{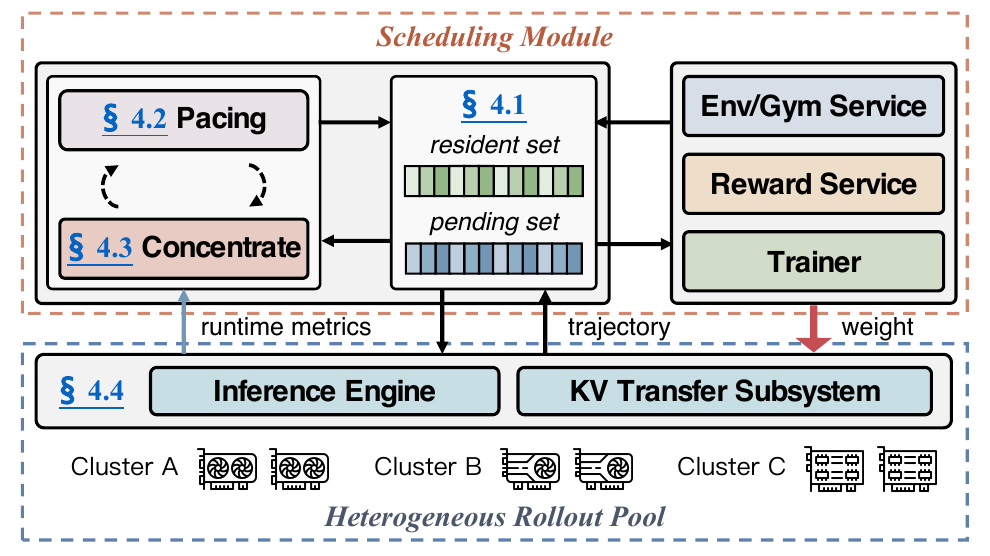}
  \caption{System architecture of \sysname{}.}
  \label{fig:design:architecture}
\end{figure}

The scheduling objective identified
in \S\ref{sec:principles}
is specialization:
letting throughput workers
sustain large active batches
while tail workers
concentrate decode resources
on long-context trajectories.
\sysname{} does not assign these roles.
The scheduler continuously reshapes
the pool's workload distribution,
and the throughput/tail distinction
emerges from the resulting workload compositions.
Two mechanisms adjust the throughput/tail worker ratio
under different conditions.
Pacing operates while fresh work remains available,
and concentration takes over when it is exhausted.
The two mechanisms share a control structure
in which trajectories move between workers
through an intermediate pending state.
Figure~\ref{fig:design:architecture}
shows the overall architecture.

\subsection{Separating Departure from Placement}
\label{sec:design:control}

\heading{A shared workload space.}
Existing RL post-training frameworks 
assign each trajectory
to a worker at dispatch time.
The assignment is fixed during execution
and changes only at event boundaries
such as weight updates
or version transitions~\cite{sheng2024hybridflow,fu2026areal,wang2025reinforcement,slime_github}.
\sysname{} instead treats the rollout pool
as a shared workload space:
trajectories can leave one worker
and resume on another as the iteration progresses.
Each worker $w$ maintains a \emph{resident set}
$\mathcal{A}_w$ of its currently assigned trajectories.
The backend engine at $w$
makes local execution decisions---%
token batching~\cite{yu2022orca}, 
memory management~\cite{kwon2023efficient}, 
and prefill ordering~\cite{agrawal2024taming}---%
while \sysname{} controls only
membership in $\mathcal{A}_w$.

\heading{The pending set.}
A trajectory not resident on any worker
belongs to a \emph{pending set} $\mathcal{P}_v$,
indexed by the policy version $v$
under which it was generated.
The version index ensures
that a trajectory resumes only on a worker
serving the same policy version.
A trajectory's lifecycle therefore extends
beyond the conventional
dispatch $\to$ execute $\to$ complete:
it cycles between active decode
and the pending state
as the scheduler reshapes the pool's workload.
$\mathcal{P}_v$ plays two roles
in the scheduling strategy.
First, it is the \emph{transit station}
for workload transformation:
when the scheduler displaces a long-context resident
and shorter-context work takes its place,
both pass through $\mathcal{P}_v$.
Second, it provides \emph{backpressure}:
a pending trajectory is not actively decoding,
so growth of $\mathcal{P}_v$ signals
that evictions are outpacing placements.
\S\ref{sec:design:tail} develops
how this backpressure
governs the pace of consolidation.

\heading{KV headroom.}
Each worker $w$ has a KV-cache capacity
determined by available accelerator memory.
\emph{KV headroom} $H_w$ is the capacity remaining
after accounting for all resident KV state
and the required reserve.
Headroom serves three roles.
For \emph{feasibility}, 
a trajectory whose footprint exceeds $H_w$ 
cannot be placed.
For \emph{composition}, 
longer-context residents consume more capacity.
For \emph{progress}, 
as residents grow and complete at different rates, 
headroom tracks the capacity each worker exposes.
Headroom is the shared state
that connects departure to placement:
an eviction raises it immediately on the source;
a placement lowers it immediately on the destination.

\heading{Departure without destination.}
A departure moves a trajectory
from $\mathcal{A}_w$ to $\mathcal{P}_v$.
Departures arise from scheduler-driven eviction
or from natural pause points
such as agentic turn boundaries
while a trajectory waits
for an environment response.
The direction of eviction---%
which trajectory to remove---%
is determined by the workload transformation
the scheduler is pursuing.
The destination depends on the pool state
at the time of placement,
which may have changed since departure.
\sysname{} therefore defers destination binding:
a departure commits to a direction,
not a target.
This separation lets both scheduling mechanisms
(\S\ref{sec:design:pacing}, \S\ref{sec:design:tail})
share the same control interface
while using different eviction
and placement directions.

\subsection{Sustaining Specialization}
\label{sec:design:pacing}

While fresh work remains available,
the scheduler realizes
the \emph{long-out, short-in} transformation
through a repeating scheduling cycle.
 
\heading{Headroom as a scheduling signal.}
The entire cycle relies on KV headroom alone.
The scheduler reads this single number
for every decision.
Headroom below $H_{\min}$ triggers eviction,
and the smallest post-placement headroom
identifies the tightest feasible destination
for each pending trajectory.
No hardware-type labels,
workload classifiers,
or role assignments
enter the scheduling loop.
 
\begin{figure}[t]
  \centering
  \includegraphics[width=\columnwidth]{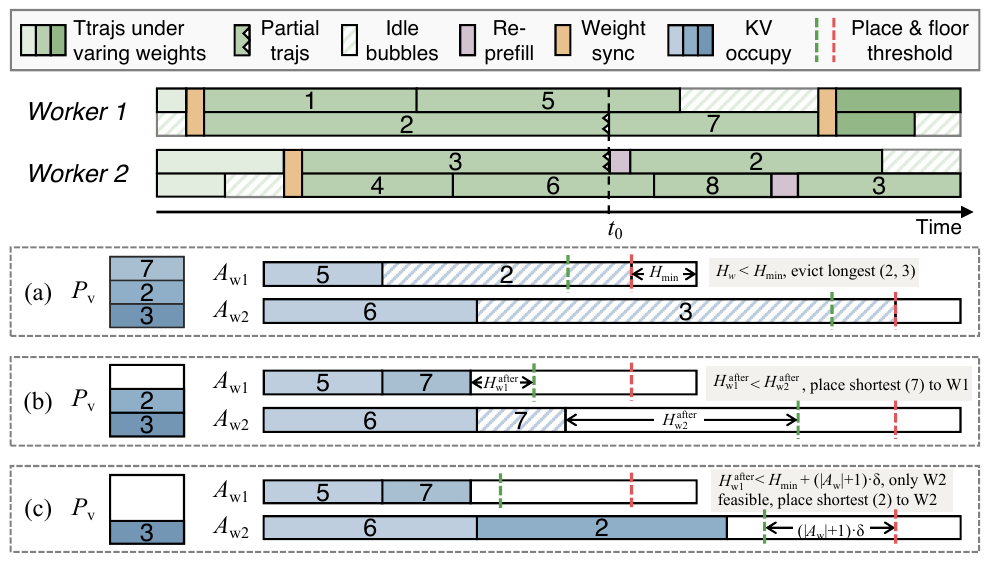}
  \caption{One simplified scheduling cycle.
  \textbf{Top:} execution timeline of two rollout workers.
  \textbf{Bottom:} the scheduler observes the pool state at time $t_0$.}
  \label{fig:design:scheduling-cycle}
\end{figure}

\heading{The scheduling cycle.}
Each cycle restores headroom on constrained workers,
then places pending work
into the resulting gaps.
When a worker's headroom falls below $H_{\min}$,
the scheduler evicts its longest resident
into $\mathcal{P}_v$, as shown for Worker~1 in 
Figure~\hyperref[fig:design:scheduling-cycle]{%
  \ref*{fig:design:scheduling-cycle}(a)%
}.
The longest resident releases the most KV capacity
per departure
and removes the greatest context pressure.
If one departure is insufficient,
eviction repeats while eligible residents remain.
The floor is restored before new work is admitted.
Pending trajectories in $\mathcal{P}_v$
are processed in ascending order of context length.
Shorter trajectories fit on more workers,
so processing them first preserves
the larger capacity gaps
for longer trajectories
that have fewer feasible destinations.
For each pending trajectory $r$,
the scheduler selects the feasible worker
with the smallest post-placement headroom:
\begin{equation}
\begin{aligned}
w^*
&=
\arg\min_{w \in \mathcal{F}(r)}
H_w^{\mathrm{after}}(r), \\
\mathcal{F}(r)
&=
\left\{
w \in \mathcal{W} :
H_w^{\mathrm{after}}(r) \geq H_{\min}
+ (|\mathcal{A}_w|{+}1)\cdot\delta
\right\}.
\end{aligned}
\end{equation}
Tight fit fills the smallest gap first
and preserves workers with larger headroom
for trajectories that require more capacity.
In Figure~\hyperref[fig:design:scheduling-cycle]{\ref*{fig:design:scheduling-cycle}(b)},
Trajectory~7 fits on both workers
but is placed on Worker~1,
whose post-placement headroom is smaller.
In~\hyperref[fig:design:scheduling-cycle]{(c)},
Trajectory~2 cannot fit on Worker~1
because its placement
would violate the placement threshold;
it is directed to Worker~2 instead.
Trajectory~3 finds no feasible worker
and remains in $\mathcal{P}_v$
until a completion or a later eviction
releases sufficient capacity.
 
\heading{From tight fit to specialization.}
Without tight fit,
short pending trajectories
would spread across all feasible workers,
and every worker would maintain
a similar workload composition,
so no specialization would emerge.
Tight fit breaks this uniformity.
Its $\arg\min$ rule
systematically directs short work
to workers with low headroom
and reserves workers with high headroom
for long trajectories
that have no tighter destination.
This creates a positive feedback loop.
A worker restored to slightly above $H_{\min}$
is selected by tight fit,
receives short work,
and remains nearly full,
making it a throughput worker.
A worker with substantial headroom
is passed over for short work,
accepts long trajectories instead,
and accumulates long-context residents,
making it a tail worker.
Over successive cycles,
the feedback amplifies
any initial headroom difference
into persistent role differentiation.
In a homogeneous pool,
these initial differences are stochastic
because different workers happen to have
different completion timings,
and tight fit amplifies them
into workload-driven specialization.
 
\heading{Natural load balancing.}
In a heterogeneous pool,
processing speed differences
interact with the eviction--replacement cycle
to match work difficulty to hardware capability.
Faster hardware decodes more tokens per unit time,
so its residents' contexts grow faster.
When $H_{\min}$ triggers,
the evicted longest resident
has grown to a longer context,
freeing more capacity.
Tight fit then routes
the shortest pending trajectories
to tighter fits elsewhere,
so the faster worker receives
the relatively longer ones
that remain.
Over successive cycles,
this shifts the faster worker's composition
toward longer average contexts,
and the shift is self-reinforcing
because longer compositions
produce even longer evictions
and even more freed capacity.
Faster hardware therefore handles
a wider range of context lengths,
including the longer trajectories
that benefit most from fast decoding.
Slower hardware handles
a narrower, shorter range.
Because harder work flows to more capable hardware,
the effective processing rates
across hardware types converge
despite their raw speed differences,
allowing the pool to sustain balanced progress
without any explicit load-balancing logic.
 
\heading{Pacing thresholds.}
Two headroom thresholds govern the cycle.
A fixed \emph{headroom floor} $H_{\min}$
defines the minimum headroom every worker must maintain.
Falling below $H_{\min}$ triggers eviction.
A higher \emph{placement threshold}
adds a per-resident \emph{decode reserve} $\delta$
that absorbs the KV growth of ongoing decoding:
a pending trajectory $r$ may be placed on worker $w$
only when
\begin{equation}
H_w^{\mathrm{after}}(r)
\;\geq\;
H_{\min} + (|\mathcal{A}_w|{+}1)\cdot\delta,
\end{equation}
where $H_w^{\mathrm{after}}(r)$ is the headroom
remaining after placing $r$
and the factor $(|\mathcal{A}_w|{+}1)$
includes the incoming trajectory.
The gap between the two thresholds
stabilizes the feedback loop.
An eviction may restore headroom above $H_{\min}$
while the worker remains below the placement threshold,
so it does not immediately accept work
that would push it back toward the floor.
The decode reserve~$\delta$
also governs pacing aggressiveness.
A larger~$\delta$ raises the placement threshold,
making replacement less frequent
but leaving more room
for ongoing context growth.
A smaller~$\delta$ allows more aggressive reshaping
at the cost of more frequent headroom pressure.
Together with the staleness-driven admission budget,
$\delta$ forms one of the two tunable parameters
that control the throughput--latency balance.

\subsection{Concentrating the Residual Tail}
\label{sec:design:tail}

The throughput--latency trade-off
in \S\ref{sec:principles:tradeoff}
reaches its turning point
when the admission budget
for the current policy version is exhausted.
No further fresh work may enter the system,
and the remaining in-flight trajectories
must drain to completion.
The scheduler shifts its objective
from sustaining throughput through pacing
to accelerating the drain
of the residual long-context tail
by concentrating it onto the workers
best equipped to serve it
and freeing weaker workers
for the next policy version.
This is a mode switch
from passively emerging specialization
to actively creating tail workers.

\heading{The pending set across phases.}
During pacing,
$\mathcal{P}_v$ acts as a transit station
with balanced flow:
long-context trajectories enter
while shorter-context trajectories leave
for workers with freed capacity.
During concentration,
$\mathcal{P}_v$ becomes a pace governor.
Displacement from evacuation sources
outpaces absorption by placement targets,
and the resulting growth of $\mathcal{P}_v$
throttles the displacement rate.
Both phases use the same control interface,
with departures entering $\mathcal{P}_v$,
placements leaving $\mathcal{P}_v$,
and headroom serving as the connecting signal.
Pacing exploits
the \emph{direction} of flow through $\mathcal{P}_v$,
while concentration exploits
the \emph{feedback} from its accumulation.
This self-regulation avoids
predicting absorption capacity.
In multi-turn settings,
trajectory completion times
depend on turn counts
and environment latencies
whose predictions are unreliable.
The accumulation of $\mathcal{P}_v$
replaces this estimate
with observable feedback.

\heading{Affinity rank and placement threshold.}
The scheduler introduces
an \emph{affinity rank} $\rho(w)$
based on each hardware type's
comparative advantage
across the workload shapes
involved in concentration.
The rank is ordinal.
Estimating per-placement benefit
would require predicting
each trajectory's remaining demand,
which is the circular dependency
identified in \S\ref{sec:principles:circular}.
An ordinal rank suffices
because each hardware type's relative strength
is stable during execution.
Placement during concentration
uses only the headroom floor $H_{\min}$,
omitting the decode reserve $\delta$
because the trajectories being consolidated
are near the end of their lifetime.
The growth budget is therefore less necessary,
and tighter packing on high-affinity workers
is the objective.

\heading{Gradual tail consolidation.}
Consolidation proceeds one source at a time
rather than reassigning the entire residual tail
in a single pass.
The scheduler selects
a single \emph{advancing source}---%
the worker with the weakest long-context affinity
among those still serving policy version $v$:
\begin{equation}
\mathit{source}
=
\arg\min_{w \in \mathcal{W}_v}
\left(
\rho(w),\,
|\mathcal{A}_w|,\,
\sum_{r \in \mathcal{A}_w}|r|
\right).
\end{equation}
The source is departure-only:
it evicts residents into $\mathcal{P}_v$
starting with the longest
but does not accept new work.
Pending trajectories
are considered for placement
in descending order of context length.
Candidate targets are ordered
by decreasing affinity rank,
then by decreasing headroom margin.
A pending trajectory $r$
may be placed on target $w$ when
\begin{equation}
H_w^{\mathrm{after}}(r) \;\geq\; H_{\min}.
\end{equation}
If a target lacks sufficient headroom,
the scheduler evicts the target's shortest residents
until the incoming long trajectory fits
or no eligible residents remain.
Displaced short trajectories enter $\mathcal{P}_v$
for a later placement pass;
they are not returned to the source,
because the source must become empty
before it can advance.

\heading{Advancing across policy versions.}
The source advances
once its resident set is empty.
It switches to the next policy version
and receives fresh trajectories
that begin at short context lengths,
restoring the conditions for pacing
on the advancing worker.
Other workers may continue serving version $v$,
draining the long trajectories
concentrated during the consolidation phase.

\subsection{KV Preparation for Late-Bound Placement}
\label{sec:design:kv}

\begin{figure}[t]
  \centering
  \includegraphics[width=\columnwidth]{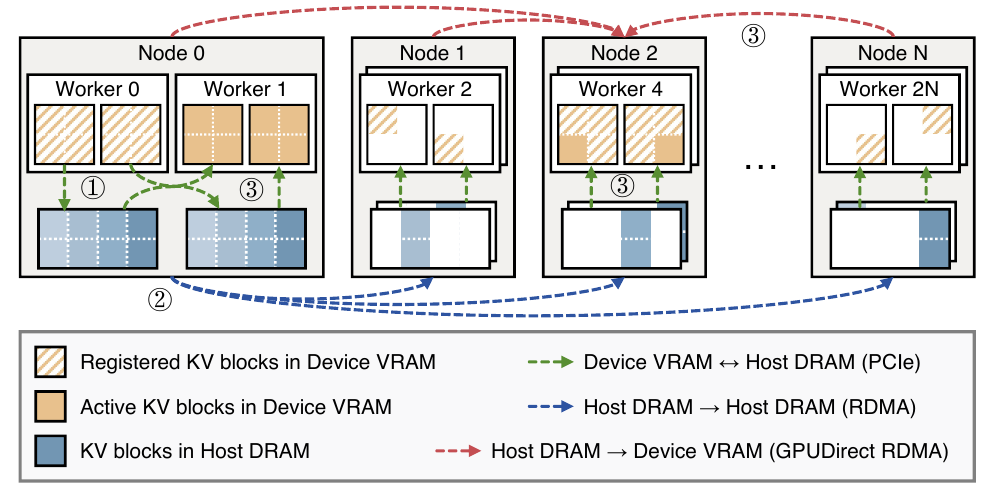}
  \caption{Three-phase KV preparation pipeline.
  \ding{192}~The source snapshots KV blocks
  from device VRAM to host DRAM over PCIe.
  \ding{193}~Prefix ranges are staged
  to other nodes via RDMA
  before a destination is selected.
  \ding{194}~After late binding,
  the target loads KV state in parallel
  from either local host DRAM via PCIe
  or remote host DRAM via GPUDirect RDMA.}
  \label{fig:design:kv-preparation}
\end{figure}

The scheduling mechanisms above
defer destination binding:
a trajectory departs
without knowing where it will resume.
A trajectory may also leave accelerator memory
at a natural pause point,
such as an agentic turn boundary
while it waits for an environment response.
When a later placement assigns the trajectory
to a different worker,
the accumulated KV prefix
must be reconstructed on the destination.
Without preparation,
this requires re-prefilling
the entire context on the new worker~%
\cite{qin2025mooncake,patel2024splitwise,zhong2024distserve}.

\sysname{} avoids this cost
through a three-phase preparation pipeline
that proceeds independently
of the scheduler's placement decisions~\cite{zhou2026tensorcast},
as shown in Figure~\ref{fig:design:kv-preparation}.
In phase~\ding{192},
the source snapshots KV state
to host-side storage
when the trajectory departs,
freeing HBM while preserving the source
as a supplier.
In phase~\ding{193},
selected prefix ranges are staged
to relay nodes over RDMA
before a destination is bound;
relay copies are additive
and staging proceeds concurrently
with scheduling decisions.
In phase~\ding{194},
the target constructs a load plan
that maps each prefix range to a supplier
and materializes the full prefix
through parallel fan-in.
Prepared transfer is used
between compatible workers
within the same cluster;
cross-cluster or cross-type migrations
fall back to token transfer
with destination-side re-prefill.

\section{Implementation}
\label{sec:implement}

The prototype is built on
vLLM~\cite{kwon2023efficient} 
as the inference backend
and Steptron~\cite{steptron_oss_github}
as the training backend.
\sysname{}'s control plane
manages resident-set membership
and the pending set at the framework level;
each vLLM instance retains its local scheduler
for token batching and memory management
but does not make cross-worker placement decisions.

\heading{Scheduling parameters.}
The scheduling policies
in \S\ref{sec:design}
are governed by two parameters:
the decode reserve $\delta$
and an admission budget that constrains staleness.
The decode reserve $\delta$ is configured in tokens
and converted to KV-cache space
using the model's per-token KV footprint.
Following Slime~\cite{slime_github},
the admission budget uses a credit model
rather than a per-trajectory cap.
Each completed training iteration
adds one iteration-batch of samples to the credit pool,
and a dimensionless parameter $\eta$
controls how far ahead of training consumption 
the system may admit new work.
In steady state,
the number of in-flight samples
is bounded by approximately
$(\eta{+}1)$ iteration-batches,
so $\eta{=}1$ corresponds
to one-step off-policy execution.
The budget controls the rate of admission,
not the age of individual trajectories.

\section{Evaluation}
\label{sec:evaluation}

\begin{figure*}[t]
  \centering
  \includegraphics[width=\textwidth]{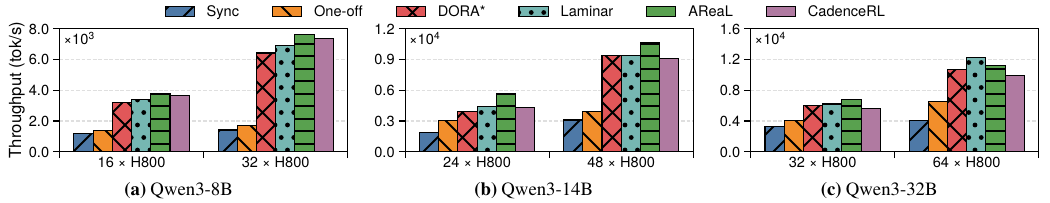}
  \caption{Aggregate decode throughput on homogeneous H800 pools across three model sizes.
  The $x$-axis denotes the total device count (trainer and rollout combined);
  the trainer is provisioned with spare capacity so that rollout is the throughput bottleneck and 
  differences between strategies reflect rollout scheduling efficiency.}
  \label{fig:evaluation:throughput}
\end{figure*}

\begin{figure*}[t]
  \centering
  \includegraphics[width=\textwidth]{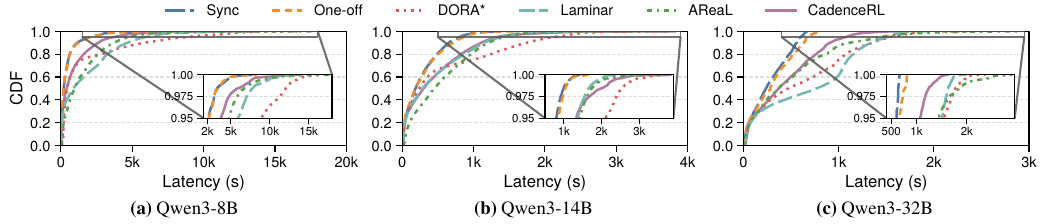}
  \caption{Cumulative distributions of trajectory latency on homogeneous H800 pools across three model sizes, using the larger-scale configuration in each case. 
  Insets magnify the tail of each distribution.}
  \label{fig:evaluation:trajectory-latency}
\end{figure*}

\subsection{Experimental Setup}
\label{sec:evaluation:setup}

\heading{Testbed.}
We use three types of accelerators---%
A910X, BI-V150, and H800---%
whose specifications are listed in Table~\ref{tab:background:accelerator-specs}.
Each accelerator type resides in a dedicated cluster.
Within a cluster,
devices communicate via rail-optimized RoCEv2 (8$\times$200\,Gbps RDMA);
across clusters,
a 20\,Gbps dedicated link provides connectivity.
When a single cluster's capacity is insufficient, rollout spans multiple clusters, introducing hardware heterogeneity.
Training remains on H800 throughout all experiments.

\heading{Datasets.}
We use a training mixture comprising datasets for mathematics reasoning~\cite{hu2026open,yu2026dapo,albalak2025big}, competitive programming~\cite{synthetic1_2025,deepcoder_2025}, STEM reasoning~\cite{fan2025megascience,rein2023gpqa,wang2023scibench}, and logical puzzles~\cite{white2025livebench,pyatkin2026generalizing,stojanovski2026reasoning}.
The mixture includes tasks with automatically verified rewards and spans single-turn generation and multi-turn environment interaction.

\heading{Models.}
We evaluate Qwen3-8B, Qwen3-14B, and Qwen3-32B~\cite{yang2025qwen3},
whose full-attention architecture makes KV-cache migration
a stress test relative to sparse or MoE variants.
We initialize RL from SFT checkpoints 
trained on the task mixture.
The inference engine's \texttt{max\_model\_len} is set to 64K, 32K, and 16K tokens for the three models respectively.

\heading{Baselines.}
We compare \sysname{} with five strategies:
Sync (sequential rollout and training, no staleness),
One-off (one-step off-policy overlap~\cite{zhong2025streamrl}),
DORA$^{*}$ (workload-aware worker reallocation~\cite{hu2026dora}),
Laminar (trajectory-level asynchrony with repack~\cite{sheng2026laminar}),
and AReaL (partial rollout with weight-switch resumption~\cite{fu2026areal}).

\heading{Settings.}
We train with PPO~\cite{schulman2017proximal} 
using a global batch size of 8192 (512 prompts, 16 responses each) and 
a single mini-batch update per training iteration.
Multi-turn tasks allow up to 12 tool calls per trajectory.
Reward and environment services run on separate CPU clusters.
Rollout workers use the minimum tensor parallelism 
that fits each model within device memory, 
maximizing the number of independent workers. 
The trainer is provisioned with spare capacity 
so that rollout generation is the throughput bottleneck.
Because the trainer never saturates, 
measured differences between strategies 
reflect rollout scheduling efficiency alone. 
All strategies share identical parallelism and device configurations.

\heading{Metrics.}
We report aggregate decode throughput (tokens\,/\,s),
trajectory latency (P50 and P95),
and tokens per dollar for cost-normalized comparison.
Trajectory latency records the elapsed time
from a trajectory's first inference step to completion,
including time spent in the pending state
between worker assignments.

\subsection{End-to-End Performance}
\label{sec:evaluation:end-to-end}

We begin with homogeneous H800 pools,
where the accelerator's maturity and
robustness to workload-composition changes (\S\ref{sec:background:hardware})
isolate the behavior of reshaping itself 
from hardware-specific effects.
We then move to heterogeneous pools, 
where accelerator complementarity 
creates additional opportunities.

\heading{Homogeneous accelerator pool.}
Figure~\ref{fig:evaluation:throughput} compares 
aggregate decode throughput on homogeneous hardware.
\sysname{} achieves throughput comparable to 
Laminar and AReaL on Qwen3-8B, 
but the gap widens at 14B and 32B.
The difference reflects two factors
of distinct character:
per-migration KV cost,
an implementation overhead
that grows with model size,
and the specialization trade-off itself.
Tail workers dedicate bandwidth to fewer trajectories
at the expense of per-worker throughput.
The first is a cost to be minimized;
the second is the mechanism
through which latency improves.
On homogeneous hardware,
neither is offset by hardware complementarity.

Figure~\ref{fig:evaluation:trajectory-latency} 
shows the gain side of this trade-off.
Sync and One-off exhibit the lowest latency
because serialized execution allocates
more memory bandwidth per trajectory
at the cost of rollout--training overlap.
Among asynchronous strategies,
\sysname{} achieves the lowest trajectory latency:
specialization keeps per-worker workload composition
controlled,
avoiding the bandwidth dilution
that arises when long and short contexts
share a worker.
Laminar and DORA$^{*}$ consolidate long-tail work
opportunistically rather than
through sustained reshaping;
AReaL mixes long and short contexts
without workload-aware placement.
Lower latency also reduces observed staleness,
as faster-completing trajectories
span fewer policy updates.
On homogeneous hardware,
specialization manifests
as an explicit throughput--latency trade-off.

\begin{figure}[t]
  \centering
  \includegraphics[width=\columnwidth]{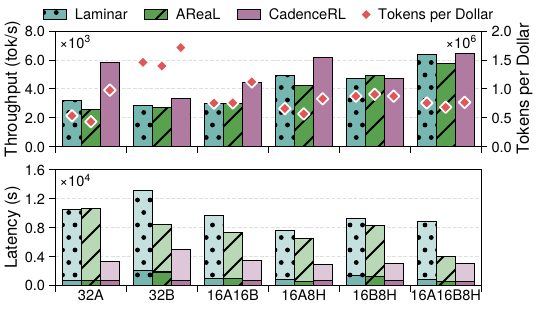}
  \caption{Aggregate decode throughput
  and P50/P95 trajectory latency
  for Qwen3-8B across rollout pool configurations.
  Pool labels indicate accelerator type
  (A\,=\,A910X, B\,=\,BI-V150, H\,=\,H800)
  and device count.
  Rollout tensor parallelism is 2 for A and B, 1 for H.
  Training remains on H800 across all runs.
  AReaL and \sysname{} both use
  admission budget $\eta{=}1$;
  \sysname{} uses
  decode reserve $\delta{=}512$.}
  \label{fig:evaluation:hetero}
\end{figure}

\heading{Heterogeneous accelerator pool.}
Figure~\ref{fig:evaluation:hetero} compares Laminar, AReaL, and \sysname{}%
\footnote{DORA$^{*}$'s reallocation model assumes homogeneous hardware;
we exclude it from heterogeneous comparisons.}
across heterogeneous pool configurations.
\sysname{} configures the affinity rank
as $\rho(\text{H800}) > \rho(\text{BI-V150}) > \rho(\text{A910X})$,
reflecting each hardware type's comparative advantage
at the workload shapes involved in concentration.
Hardware complementarity
transforms the trade-off visible on homogeneous pools
into a synergy:
\sysname{} matches or exceeds baseline throughput
across all configurations
while substantially reducing trajectory latency.
High-bandwidth accelerators (H800)
serve as natural tail workers
because their bandwidth advantage
accelerates long-context decode
without sacrificing proportional throughput;
cost-efficient accelerators
sustain large active batches
within their effective operating range.
\sysname{}'s P95 latency remains nearly constant
across H800-containing configurations,
confirming that the long-context tail
is consistently routed
to the accelerator best equipped to drain it.
In the 16A16B pool (no H800),
\sysname{} achieves the highest tokens per dollar
by directing each accelerator type
toward workloads it serves most efficiently.
Adding H800 improves absolute throughput and latency
but dilutes cost efficiency
due to its higher per-device price.
\sysname{} makes heterogeneous scaling composable:
adding high-bandwidth accelerators
reduces tail latency,
while adding cost-efficient accelerators
increases throughput,
with no manual routing configuration required.

\begin{figure}[t]
  \centering
  \includegraphics[width=\columnwidth]{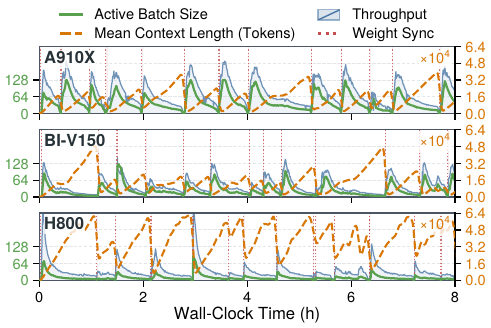}
  \caption{Runtime workload dynamics on 
  a heterogeneous pool (A910X, BI-V150, H800) under \sysname{}.
  Throughput is plotted over a 0--2000~tokens\,/\,s range.}
  \label{fig:evaluation:overtime}
\end{figure}

\heading{Runtime workload dynamics.}
Figure~\ref{fig:evaluation:overtime} traces runtime workload dynamics on a heterogeneous pool over multiple training iterations.
H800 workers sustain substantially higher mean context lengths than workers on A910X and BI-V150 throughout the run,
confirming that long-context work is consistently routed to the accelerator with the strongest long-context affinity.
Specialization is not rigid. 
Under high load, 
all accelerator types 
carry elevated context lengths; 
when load subsides, 
specialization re-emerges.

\begin{figure}[!tb]
  \centering
  \includegraphics[width=\columnwidth]{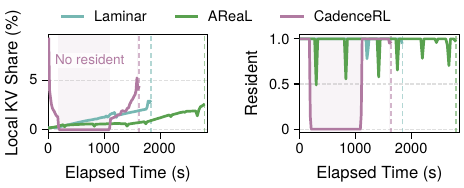}
  \caption{Lifecycle of a long-context trajectory under three asynchronous strategies on homogeneous Qwen3-32B configuration.
  \emph{Local KV share}: the trajectory's KV-cache occupancy divided by the host worker's total KV-cache occupancy.
  \emph{Resident}: whether the trajectory is actively being decoded.
  Vertical dashed lines mark trajectory completion.}
  \label{fig:evaluation:tail-latency-analysis}
\end{figure}

\heading{Long-tail trajectory behavior.}
To understand why \sysname{} reduces tail latency, 
we trace a single long-context trajectory 
(approximately 16K generated tokens) 
under each asynchronous strategy
on the homogeneous Qwen3-32B configuration,
measuring its local KV share and residency status
over its lifecycle.
Figure~\ref{fig:evaluation:tail-latency-analysis} shows both quantities.
Under \sysname{}, the trajectory initially receives a high local KV share during the pacing phase,
when it is short and well-matched to its host worker.
As the context grows through continued decoding, the trajectory outgrows the worker's effective capacity and 
is yielded to the pending pool.
It re-enters on a worker with sufficient headroom,
again at high KV share.
This yield-and-return cycle repeats:
the trajectory is resident for less than half of its lifecycle
but concentrates decode into high-share windows where per-token progress is fast.
Laminar keeps the trajectory resident continuously at a persistently low KV share,
making slow but steady progress throughout.
AReaL also maintains near-continuous residency but repeatedly interrupts execution with partial rollouts,
each time resuming at a low KV share.
Despite spending the least time in active decode,
\sysname{}'s trajectory completes first---%
demonstrating that workload specialization translates directly into tail-latency reduction.

\subsection{Sensitivity Analysis}
\label{sec:evaluation:sensitivity}

\sysname{} exposes two scheduling parameters:
the admission budget $\eta$ and the decode reserve $\delta$.
We vary each independently and verify that the resulting configurations preserve training quality.

\begin{figure}[!tb]
  \centering
  \includegraphics[width=\columnwidth]{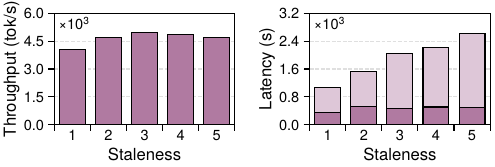}
  \caption{Decode throughput and P50/P95 trajectory latency sensitivity to staleness $\eta$ on Qwen3-32B.}
  \label{fig:evaluation:staleness}
\end{figure}

\heading{Staleness--performance trade-off.}
Figure~\ref{fig:evaluation:staleness} varies the staleness parameter $\eta$ from 1 to 5 on Qwen3-32B.
Increasing $\eta$ from 1 to 3 improves throughput by allowing more rollout--training overlap.
Beyond $\eta{=}3$, throughput declines.
Each policy version's supply of fresh work is fixed;
beyond the point where pacing already consumes it fully,
additional overlap only prolongs concentration
rather than increasing throughput.
Latency increases monotonically with $\eta$, 
due to more concurrent in-flight trajectories.

\begin{figure}[!tb]
  \centering
  \includegraphics[width=\columnwidth]{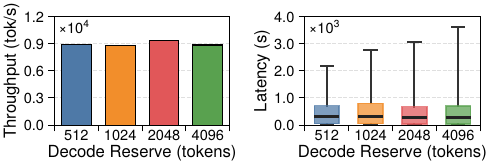}
  \caption{Decode throughput and trajectory latency sensitivity to decode reserve $\delta$ on Qwen3-14B.}
  \label{fig:evaluation:reserve}
\end{figure}

\heading{Decode-reserve sensitivity.}
The decode reserve $\delta$, specified in tokens, reserves KV-cache space per trajectory to absorb future token growth.
Figure~\ref{fig:evaluation:reserve} varies $\delta$ from 512 to 4096 tokens on Qwen3-14B.
Aggregate throughput remains stable across all settings; the effect appears in the tail, where P95 latency rises substantially as $\delta$ increases.
A larger reserve reduces the headroom available for placement, limiting scheduling freedom and weakening tail-latency suppression.
 
\begin{figure}[!tb]
  \centering
  \includegraphics[width=\columnwidth]{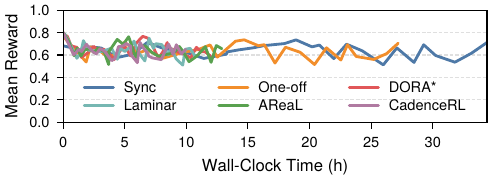}
  \caption{Mean reward over 30 training iterations for six rollout scheduling strategies on Qwen3-8B.
  Asynchronous strategies use one-step staleness ($\eta{=}1$).}
  \label{fig:evaluation:train-stability}
\end{figure}

\heading{Training quality}
Figure~\ref{fig:evaluation:train-stability} shows
that reward curves closely overlap
across all six strategies under $\eta{=}1$,
confirming that \sysname{}
does not measurably affect training quality.

\subsection{KV Cache Transfer Overhead}
\label{sec:evaluation:kv-transfer}

\begin{figure}[!tb]
  \centering
  \includegraphics[width=\columnwidth]{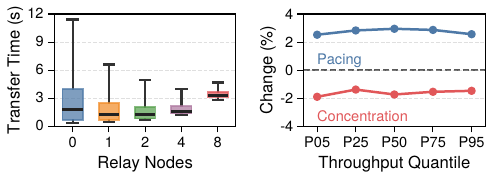}
  \caption{KV cache transfer overhead on Qwen3-8B. 
  \textbf{Left:} critical-path transfer time across relay node counts. 
  \textbf{Right:} decode throughput change under 2-relay-node KV preparation relative to no preparation (dashed line), by scheduling phase and throughput quantile.}
  \label{fig:evaluation:kv-transfer}
\end{figure}

Figure~\ref{fig:evaluation:kv-transfer} 
quantifies the overhead of KV cache preparation on Qwen3-8B.
The P50 critical-path transfer time stabilizes
with a small number of relay nodes.
The tail continues to shrink as more relay nodes are added,
but with diminishing returns.
The throughput impact depends on the scheduling phase.
During pacing, the active batch is large,
and KV preparation yields a modest throughput gain.
During concentration, fewer trajectories are active,
and the transfer overhead slightly reduces throughput.
In both phases, the effect remains within a few percent
but has opposite signs.
This suggests that KV preparation should be configured
based on workload characteristics
rather than applied uniformly.

\section{Related Work}
\label{sec:related}

\heading{RL execution and resource orchestration.}
RL systems vary in synchronization and resource scope~\cite{tan2026dynamic}.
AReaL~\cite{fu2026areal} interrupts generation for updates.
Laminar~\cite{sheng2026laminar} repacks trajectories around a parameter
service, while DORA~\cite{hu2026dora} streams training across
policy versions. StreamRL~\cite{zhong2025streamrl} disaggregates generation
from training and returns samples incrementally.
Weave~\cite{wu2026weave} co-schedules jobs, DynaRL~\cite{wang2026dynarl}
reallocates resources across pipeline components, HetRL~\cite{he2026hetrl}
searches heterogeneous execution plans, and JigsawRL~\cite{hu2026jigsawrl}
multiplexes pipeline sub-stages. These systems operate at pipeline/resource
scope. \sysname{} reshapes unfinished generations within one
rollout pool.

\heading{Trajectory execution and tail scheduling.}
Seer~\cite{qin2026seer} divides prompt groups into generation chunks and
uses global KV with speculative length estimates.
Heddle~\cite{zhang2026heddle} predicts trajectory progress for priority
scheduling, migration, and model-parallel specialization.
APRIL~\cite{zhou2025april} carries unfinished oversampled responses across
iterations. TailSieve~\cite{xu2026tailsieve} uses partial rollouts only as
tail-routing signals and then regenerates on-policy. These methods act on
chunks, iterations, or routing pools. \sysname{} instead controls worker
resident sets. Its pending set $\mathcal{P}_v$ separates departure from
placement, while KV headroom drives pacing and concentration without fixed
chunks or per-migration benefit prediction.

\heading{Serving and KV-cache management.}
Serving systems provide selective batching~\cite{yu2022orca}, paged KV
allocation~\cite{kwon2023efficient}, chunked prefill~\cite{agrawal2024taming},
disaggregated prefill/decode~\cite{patel2024splitwise,zhong2024distserve},
and distributed KV pools~\cite{qin2025mooncake,zhou2026tensorcast}. 
MISA-T~\cite{hong2026scheduling}
adds workload-aware session admission and residency-time-aware KV accounting
while preserving the trainer's workload mixture. \sysname{} is
complementary. It changes each policy-version rollout pool's composition,
stages KV state for late-bound placement, and falls back to re-prefill when
state cannot be reused.

\section{Conclusion}
\label{sec:conclusion}

We presented \sysname{},
a scheduling system
for heterogeneous rollout pools
in RL post-training.
The central insight is that
high throughput and fast tail completion
require different workload compositions
rather than a per-worker compromise,
and the decode-throughput landscape
structurally rewards reshaping toward this specialization,
thereby bypassing per-migration benefit estimation.
Two mechanisms---pacing and concentration---%
realize this reshaping
through a shared control structure
governed by two parameters.
On heterogeneous hardware,
complementary accelerator strengths
transform the homogeneous-pool trade-off into a synergy:
scaling is composable,
with different accelerator types
improving different performance axes
without manual routing.


\bibliographystyle{plain}
\bibliography{references}

@article{sheng2024hybridflow,
  title={Hybridflow: A flexible and efficient rlhf framework},
  author={Sheng, Guangming and Zhang, Chi and Ye, Zilingfeng and Wu, Xibin and Zhang, Wang and Zhang, Ru and Peng, Yanghua and Lin, Haibin and Wu, Chuan},
  journal={arXiv preprint arXiv:2409.19256},
  year={2024}
}

@article{fu2026areal,
  title={Areal: A large-scale asynchronous reinforcement learning system for language reasoning},
  author={Fu, Wei and Gao, Jiaxuan and Shen, Xujie and Zhu, Chen and Mei, Zhiyu and He, Chuyi and Xu, Shusheng and Wei, Guo and Mei, Jun and Wang, Jiashu and others},
  journal={Advances in Neural Information Processing Systems},
  volume={38},
  pages={36256--36282},
  year={2026}
}

@article{wang2025reinforcement,
  title={Reinforcement Learning Optimization for Large-Scale Learning: An Efficient and User-Friendly Scaling Library},
  author={Wang, Weixun and Xiong, Shaopan and Chen, Gengru and Gao, Wei and Guo, Sheng and He, Yancheng and Huang, Ju and Liu, Jiaheng and Li, Zhendong and Li, Xiaoyang and others},
  journal={arXiv preprint arXiv:2506.06122},
  year={2025}
}

@misc{slime_github,
  author       = {Zilin Zhu and Chengxing Xie and Xin Lv and slime Contributors},
  title        = {slime: An LLM post-training framework for RL Scaling},
  year         = {2025},
  howpublished = {\url{https://github.com/THUDM/slime}},
  urldate      = {2025-06-19}
}

@inproceedings{wang2026dynarl,
  title={{DynaRL}: Flexible and Dynamic Scheduling of {Large-Scale} Reinforcement Learning Training},
  author={Wang, Yuanqing and Lin, Hao and Hu, Junhao and Zhu, Chunyang and Zhang, Quanlu and Guo, Zhen and Zhang, Yuchen and Fu, Xu and Xu, Si and Dai, Bo and others},
  booktitle={20th USENIX Symposium on Operating Systems Design and Implementation (OSDI 26)},
  pages={847--862},
  year={2026}
}

@inproceedings{yu2022orca,
  title={Orca: A distributed serving system for {Transformer-Based} generative models},
  author={Yu, Gyeong-In and Jeong, Joo Seong and Kim, Geon-Woo and Kim, Soojeong and Chun, Byung-Gon},
  booktitle={16th USENIX symposium on operating systems design and implementation (OSDI 22)},
  pages={521--538},
  year={2022}
}

@inproceedings{kwon2023efficient,
  title={Efficient memory management for large language model serving with pagedattention},
  author={Kwon, Woosuk and Li, Zhuohan and Zhuang, Siyuan and Sheng, Ying and Zheng, Lianmin and Yu, Cody Hao and Gonzalez, Joseph and Zhang, Hao and Stoica, Ion},
  booktitle={Proceedings of the 29th symposium on operating systems principles},
  pages={611--626},
  year={2023}
}

@inproceedings{agrawal2024taming,
  title={Taming {Throughput-Latency} tradeoff in {LLM} inference with {Sarathi-Serve}},
  author={Agrawal, Amey and Kedia, Nitin and Panwar, Ashish and Mohan, Jayashree and Kwatra, Nipun and Gulavani, Bhargav and Tumanov, Alexey and Ramjee, Ramachandran},
  booktitle={18th USENIX symposium on operating systems design and implementation (OSDI 24)},
  pages={117--134},
  year={2024}
}

@misc{huawei_a910x,
  author       = {{Huawei Technologies}},
  title        = {Ascend NPU Computing Products},
  howpublished = {\url{https://www.hiascend.com/}},
  year         = {2026}
}

@misc{iluvatar_biv150,
  author       = {{Iluvatar CoreX}},
  title        = {Iluvatar CoreX GPU Computing Products},
  howpublished = {\url{https://iluvatar.com/}},
  year         = {2026}
}

@misc{nvidia_h800,
  author       = {{NVIDIA}},
  title        = {NVIDIA GPU Computing Products},
  howpublished = {\url{https://www.nvidia.com}},
  year         = {2026}
}

@article{yao2022react,
  title={React: Synergizing reasoning and acting in language models},
  author={Yao, Shunyu and Zhao, Jeffrey and Yu, Dian and Du, Nan and Shafran, Izhak and Narasimhan, Karthik and Cao, Yuan},
  journal={arXiv preprint arXiv:2210.03629},
  year={2022}
}

@article{wang2024survey,
  title={A survey on large language model based autonomous agents},
  author={Wang, Lei and Ma, Chen and Feng, Xueyang and Zhang, Zeyu and Yang, Hao and Zhang, Jingsen and Chen, Zhiyuan and Tang, Jiakai and Chen, Xu and Lin, Yankai and others},
  journal={Frontiers of computer science},
  volume={18},
  number={6},
  pages={186345},
  year={2024},
  publisher={Springer}
}

@article{shao2024deepseekmath,
  title={Deepseekmath: Pushing the limits of mathematical reasoning in open language models},
  author={Shao, Zhihong and Wang, Peiyi and Zhu, Qihao and Xu, Runxin and Song, Junxiao and Bi, Xiao and Zhang, Haowei and Zhang, Mingchuan and Li, YK and Wu, Yang and others},
  journal={arXiv preprint arXiv:2402.03300},
  year={2024}
}

@article{guo2025deepseek,
  title={Deepseek-r1: Incentivizing reasoning capability in llms via reinforcement learning},
  author={Guo, Daya and Yang, Dejian and Zhang, Haowei and Song, Junxiao and Wang, Peiyi and Zhu, Qihao and Xu, Runxin and Zhang, Ruoyu and Ma, Shirong and Bi, Xiao and others},
  journal={arXiv preprint arXiv:2501.12948},
  year={2025}
}

@inproceedings{jimenez2024swe,
  title={Swe-bench: Can language models resolve real-world github issues?},
  author={Jimenez, Carlos E and Yang, John and Wettig, Alexander and Yao, Shunyu and Pei, Kexin and Press, Ofir and Narasimhan, Karthik},
  booktitle={International Conference on Learning Representations},
  volume={2024},
  pages={54107--54157},
  year={2024}
}

@article{rein2023gpqa,
  title={Gpqa: A graduate-level google-proof q\&a benchmark},
  author={Rein, David and Hou, Betty Li and Stickland, Asa Cooper and Petty, Jackson and Pang, Richard Yuanzhe and Dirani, Julien and Michael, Julian and Bowman, Samuel R},
  journal={arXiv preprint arXiv:2311.12022},
  year={2023}
}

@article{hendrycks2020measuring,
  title={Measuring massive multitask language understanding},
  author={Hendrycks, Dan and Burns, Collin and Basart, Steven and Zou, Andy and Mazeika, Mantas and Song, Dawn and Steinhardt, Jacob},
  journal={arXiv preprint arXiv:2009.03300},
  year={2020}
}

@inproceedings{yue2024mmmu,
  title={Mmmu: A massive multi-discipline multimodal understanding and reasoning benchmark for expert agi},
  author={Yue, Xiang and Ni, Yuansheng and Zhang, Kai and Zheng, Tianyu and Liu, Ruoqi and Zhang, Ge and Stevens, Samuel and Jiang, Dongfu and Ren, Weiming and Sun, Yuxuan and others},
  booktitle={Proceedings of the IEEE/CVF conference on computer vision and pattern recognition},
  pages={9556--9567},
  year={2024}
}

@misc{meta_nvidia_2026,
  author       = {{NVIDIA Corporation}},
  title        = {Meta Builds AI Infrastructure With NVIDIA},
  howpublished = {\url{https://nvidianews.nvidia.com/news/meta-builds-ai-infrastructure-with-nvidia}},
  year         = {2026}
}

@misc{anthropic_amazon_2026,
  author       = {{Anthropic}},
  title        = {Anthropic and Amazon Expand Collaboration for up to 5 Gigawatts of New Compute},
  howpublished = {\url{https://www.anthropic.com/news/anthropic-amazon-compute}},
  year         = {2026}
}

@inproceedings{wu2026weave,
  title={Weave: Efficient {Co-Scheduling} for Disaggregated {RL} {Post-Training}},
  author={Wu, Tianyuan and Cao, Lunxi and Wei, Yining and Gao, Wei and Zhao, Yuheng and An, Dakai and Xiong, Shaopan and Lv, Zhiqiang and Huang, Ju and Yang, Siran and others},
  booktitle={20th USENIX Symposium on Operating Systems Design and Implementation (OSDI 26)},
  pages={809--827},
  year={2026}
}

@article{he2026hetrl,
  title={HetRL: Efficient reinforcement learning for LLMs in heterogeneous environments},
  author={He, Yongjun and Zhang, Shuai and Gai, Jiading and Zhang, Xiyuan and Han, Boran and Wang, Bernie and Rangwala, Huzefa and Karypis, George},
  journal={Proceedings of Machine Learning and Systems},
  volume={8},
  pages={975--996},
  year={2026}
}

@misc{steptron_oss_github,
  author       = {{StepFun AI}},
  title        = {SteptronOss},
  year         = {2026},
  howpublished = {\url{https://github.com/stepfun-ai/SteptronOss}}
}

@article{hu2026open,
  title={Open-reasoner-zero: An open source approach to scaling up reinforcement learning on the base model},
  author={Hu, Jingcheng and Zhang, Yinmin and Han, Qi and Jiang, Daxin and Zhang, Xiangyu and Shum, Heung-Yeung},
  journal={Advances in Neural Information Processing Systems},
  volume={38},
  pages={162239--162262},
  year={2026}
}

@article{yu2026dapo,
  title={Dapo: An open-source llm reinforcement learning system at scale},
  author={Yu, Qiying and Zhang, Zheng and Zhu, Ruofei and Yuan, Yufeng and Zuo, Xiaochen and Yue, Yu and Dai, Weinan and Fan, Tiantian and Liu, Gaohong and Liu, Lingjun and others},
  journal={Advances in Neural Information Processing Systems},
  volume={38},
  pages={113222--113244},
  year={2026}
}

@article{albalak2025big,
  title={Big-math: A large-scale, high-quality math dataset for reinforcement learning in language models},
  author={Albalak, Alon and Phung, Duy and Lile, Nathan and Rafailov, Rafael and Gandhi, Kanishk and Castricato, Louis and Singh, Anikait and Blagden, Chase and Xiang, Violet and Mahan, Dakota and others},
  journal={arXiv preprint arXiv:2502.17387},
  year={2025}
}

@misc{synthetic1_2025,
  author       = {Justus Mattern and Sami Jaghouar and Manveer Basra and Jannik Straube and Matthew Di Ferrante and Felix Gabriel and Jack Min Ong and Vincent Weisser and Johannes Hagemann},
  title        = {SYNTHETIC-1: Two Million Collaboratively Generated Reasoning Traces from Deepseek-R1},
  howpublished = {\url{https://www.primeintellect.ai/blog/synthetic-1-release}},
  year         = {2025}
}

@misc{deepcoder_2025,
  author       = {Michael Luo and Sijun Tan and Roy Huang and Ameen Patel and Alpay Ariyak and Qingyang Wu and Xiaoxiang Shi and Rachel Xin and Colin Cai and Maurice Weber and Ce Zhang and Li Erran Li and Raluca Ada Popa and Ion Stoica},
  title        = {{DeepCoder}: A Fully Open-Source 14B Coder at O3-mini Level},
  howpublished = {\url{https://pretty-radio-b75.notion.site/DeepCoder-A-Fully-Open-Source-14B-Coder-at-O3-mini-Level-1cf81902c14680b3bee5eb349a512a51}},
  note         = {Notion Blog},
  year         = {2025}
}

@article{fan2025megascience,
  title={Megascience: Pushing the frontiers of post-training datasets for science reasoning},
  author={Fan, Run-Ze and Wang, Zengzhi and Liu, Pengfei},
  journal={arXiv preprint arXiv:2507.16812},
  year={2025}
}

@article{wang2023scibench,
  title={Scibench: Evaluating college-level scientific problem-solving abilities of large language models},
  author={Wang, Xiaoxuan and Hu, Ziniu and Lu, Pan and Zhu, Yanqiao and Zhang, Jieyu and Subramaniam, Satyen and Loomba, Arjun R and Zhang, Shichang and Sun, Yizhou and Wang, Wei},
  journal={arXiv preprint arXiv:2307.10635},
  year={2023}
}

@inproceedings{white2025livebench,
  title={LiveBench: A challenging, contamination-limited LLM benchmark},
  author={White, Colin and Dooley, Samuel and Roberts, Manley and Pal, Arka and Feuer, Benjamin and Jain, Siddhartha and Shwartz-Ziv, Ravid and Jain, Neel and Saifullah, Khalid and Dey, Sreemanti and others},
  booktitle={International Conference on Learning Representations},
  volume={2025},
  pages={91595--91631},
  year={2025}
}

@article{pyatkin2026generalizing,
  title={Generalizing verifiable instruction following},
  author={Pyatkin, Valentina and Malik, Saumya and Graf, Victoria and Ivison, Hamish and Huang, Shengyi and Dasigi, Pradeep and Lambert, Nathan and Hajishirzi, Hanna},
  journal={Advances in Neural Information Processing Systems},
  volume={38},
  year={2026}
}

@article{stojanovski2026reasoning,
  title={Reasoning gym: Reasoning environments for reinforcement learning with verifiable rewards},
  author={Stojanovski, Zafir and Stanley, Oliver and Sharratt, Joe and Jones, Richard and Adefioye, Abdulhakeem and Kaddour, Jean and K{\"o}pf, Andreas},
  journal={Advances in Neural Information Processing Systems},
  volume={38},
  year={2026}
}

@article{zhong2025streamrl,
  title={Streamrl: Scalable, heterogeneous, and elastic rl for llms with disaggregated stream generation},
  author={Zhong, Yinmin and Zhang, Zili and Song, Xiaoniu and Hu, Hanpeng and Jin, Chao and Wu, Bingyang and Chen, Nuo and Chen, Yukun and Zhou, Yu and Wan, Changyi and others},
  journal={arXiv preprint arXiv:2504.15930},
  year={2025}
}

@article{hu2026dora,
  title={DORA: A scalable asynchronous reinforcement learning system for language model training},
  author={Hu, Tianhao and Liu, Xiangcheng and Miao, Yuchun and Xiao, Youshao and Zang, Hongyu and Zheng, Yang and Huang, Xuan and Ding, Jinrui and Zhang, Yufei and Yang, Yu and others},
  journal={arXiv preprint arXiv:2604.26256},
  year={2026}
}

@inproceedings{sheng2026laminar,
  title={Laminar: A scalable asynchronous RL post-training framework},
  author={Sheng, Guangming and Tong, Yuxuan and Wan, Borui and Zhang, Wang and Jia, Chaobo and Wu, Xibin and Wu, Yuqi and Li, Xiang and Zhang, Chi and Peng, Yanghua and others},
  booktitle={Proceedings of the 21st European Conference on Computer Systems},
  pages={400--422},
  year={2026}
}

@article{yang2025qwen3,
  title={Qwen3 technical report},
  author={Yang, An and Li, Anfeng and Yang, Baosong and Zhang, Beichen and Hui, Binyuan and Zheng, Bo and Yu, Bowen and Gao, Chang and Huang, Chengen and Lv, Chenxu and others},
  journal={arXiv preprint arXiv:2505.09388},
  year={2025}
}

@article{schulman2017proximal,
  title={Proximal policy optimization algorithms},
  author={Schulman, John and Wolski, Filip and Dhariwal, Prafulla and Radford, Alec and Klimov, Oleg},
  journal={arXiv preprint arXiv:1707.06347},
  year={2017}
}

@inproceedings{qin2026seer,
  title={Seer: Online Context Learning for Fast Synchronous {LLM} Reinforcement Learning},
  author={Qin, Ruoyu and He, Weiran and Huang, Weixiao and Zhang, Yangkun and Zhao, Yikai and Pang, Bo and Xu, Xinran and Shan, Yingdi and Wu, Yongwei and Zhang, Mingxing},
  booktitle={20th USENIX Symposium on Operating Systems Design and Implementation (OSDI 26)},
  pages={883--901},
  year={2026}
}

@article{zhou2025april,
  title={April: Active partial rollouts in reinforcement learning to tame long-tail generation},
  author={Zhou, Yuzhen and Li, Jiajun and Su, Yusheng and Ramesh, Gowtham and Zhu, Zilin and Long, Xiang and Zhao, Chenyang and Pan, Jin and Yu, Xiaodong and Wang, Ze and others},
  journal={arXiv preprint arXiv:2509.18521},
  year={2025}
}

@article{zhang2026heddle,
  title={Heddle: A Distributed Orchestration System for Agentic RL Rollout},
  author={Zhang, Zili and Zhong, Yinmin and Yang, Chengxu and Jin, Chao and Wu, Bingyang and Wei, Xinming and Liu, Yuliang and Jin, Xin},
  journal={arXiv preprint arXiv:2603.28101},
  year={2026}
}

@article{hu2026jigsawrl,
  title={JigsawRL: Assembling RL Pipelines for Efficient LLM Post-Training},
  author={Hu, Zhengding and Ouyang, Hehua and Chen, Chang and Pan, Zaifeng and Guan, Yue and Yu, Zhongkai and Wang, Zhen and Swanson, Steven and Ding, Yufei},
  journal={arXiv preprint arXiv:2604.23838},
  year={2026}
}

@article{xu2026tailsieve,
  title={TailSieve: Partial-Rollout-Guided Tail Routing for LLM Rollouts},
  author={Xu, Tianqi and Lv, Lu and Huang, Haoyang and Huang, Wenjie and Shen, Zhanming and Shen, Yuhao and Zhang, Baolin and Hu, Xinyi and Ge, Shuang and Dai, Jun and others},
  journal={arXiv preprint arXiv:2608.22788},
  year={2026}
}

@inproceedings{patel2024splitwise,
  title={Splitwise: Efficient generative llm inference using phase splitting},
  author={Patel, Pratyush and Choukse, Esha and Zhang, Chaojie and Shah, Aashaka and Goiri, {\'I}{\~n}igo and Maleki, Saeed and Bianchini, Ricardo},
  booktitle={2024 ACM/IEEE 51st Annual International Symposium on Computer Architecture (ISCA)},
  pages={118--132},
  year={2024},
  organization={IEEE}
}

@inproceedings{zhong2024distserve,
  title={{DistServe}: Disaggregating prefill and decoding for goodput-optimized large language model serving},
  author={Zhong, Yinmin and Liu, Shengyu and Chen, Junda and Hu, Jianbo and Zhu, Yibo and Liu, Xuanzhe and Jin, Xin and Zhang, Hao},
  booktitle={18th USENIX symposium on operating systems design and implementation (OSDI 24)},
  pages={193--210},
  year={2024}
}

@inproceedings{qin2025mooncake,
  title={Mooncake: Trading more storage for less computation—a {KVCache-centric} architecture for serving {LLM} chatbot},
  author={Qin, Ruoyu and Li, Zheming and He, Weiran and Cui, Jialei and Ren, Feng and Zhang, Mingxing and Wu, Yongwei and Zheng, Weimin and Xu, Xinran},
  booktitle={23rd USENIX conference on file and storage technologies (FAST 25)},
  pages={155--170},
  year={2025}
}

@article{hong2026scheduling,
  title={Scheduling Mixed RL Rollouts Beyond Prefix Locality},
  author={Hong, Zetao and Yuan, Song and Ding, Yuanhao and Zhu, Yibo and Jiang, Daxin and Wang, Zhibin and Tian, Chen},
  journal={arXiv preprint arXiv:2608.11152},
  year={2026}
}

@inproceedings{tan2026dynamic,
  title={Dynamic Compute and Network Orchestration for Disaggregated RL},
  author={Tan, Xin and Feng, Yicheng and Zhou, Yu and Jiang, Yimin and Zhu, Yibo and Xu, Hong},
  booktitle={Proceedings of the ACM SIGCOMM 2026 Conference},
  pages={1109--1126},
  year={2026}
}

@article{zhou2026tensorcast,
  title={TensorCast: The Missing Tensor Management Layer in Large Language Model Infrastructure},
  author={Zhou, Yuhan and Luo, Yuchu and Nie, Hao and Lv, Wangrunze and Zhou, Yu and Zhu, Yibo and Jiang, Daxin and Xu, Chenren},
  journal={arXiv preprint arXiv:2608.06007},
  year={2026}
}

@article{pope2023efficiently,
  title={Efficiently scaling transformer inference},
  author={Pope, Reiner and Douglas, Sholto and Chowdhery, Aakanksha and Devlin, Jacob and Bradbury, James and Heek, Jonathan and Xiao, Kefan and Agrawal, Shivani and Dean, Jeff},
  journal={Proceedings of machine learning and systems},
  volume={5},
  pages={606--624},
  year={2023}
}

@article{recasens2026slim,
  title={SLIM: Saturation-Aware Lightweight Performance Modeling for LLM Serving},
  author={Recasens, Pol G and Agullo, Ferran and Zhu, Yue and Wang, Chen and Torres, Jordi and Berral, Josep Ll},
  journal={arXiv preprint arXiv:2607.29575},
  year={2026}
}

@article{qu2025copris,
  title={Copris: Efficient and stable reinforcement learning via concurrency-controlled partial rollout with importance sampling},
  author={Qu, Zekai and Pan, Yinxu and Sun, Ao and Xiao, Chaojun and Han, Xu},
  journal={arXiv preprint arXiv:2511.05589},
  year={2025}
}


\clearpage
\appendix
\section{Rollout Coordination Mechanisms}
\label{app:coordination}
 
\begin{figure}[!ht]
  \centering
  \includegraphics[width=\columnwidth]{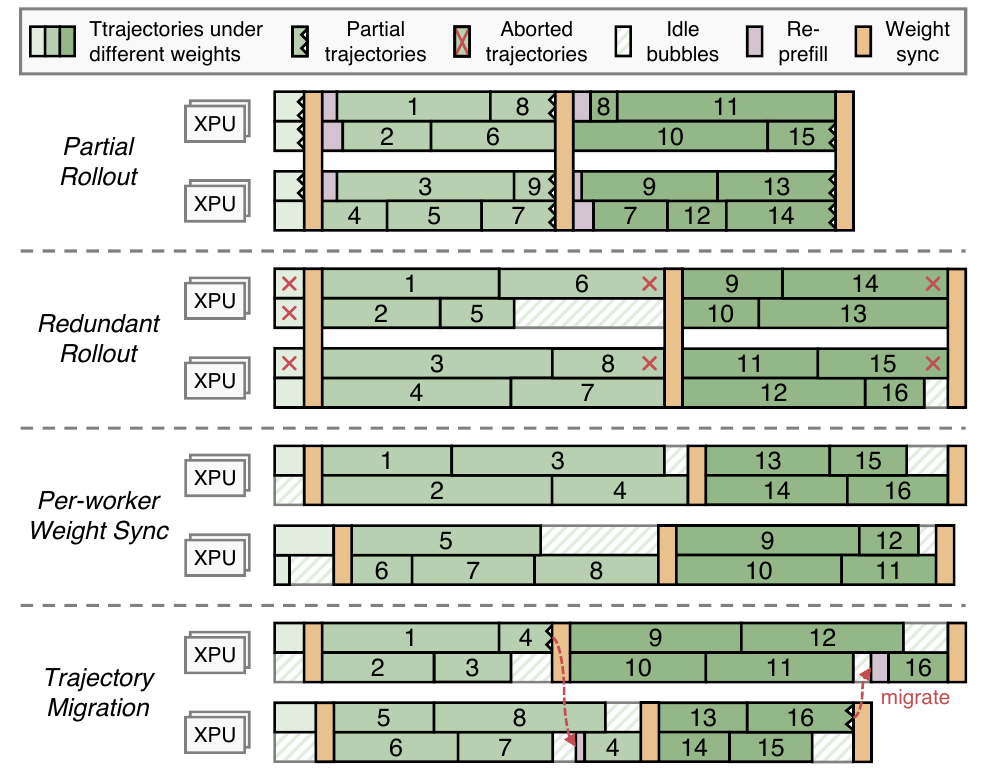}
  \caption{Rollout coordination mechanisms.}
  \label{fig:background:coordination}
\end{figure}

Existing rollout systems provide several mechanisms for handling trajectories that finish at different rates, as illustrated in Figure~\ref{fig:background:coordination}.
Partial rollout pauses unfinished trajectories and resumes them in a later iteration.
Redundant rollout launches more trajectories than training requires and discards the excess once enough have completed.
Independent weight advancement allows a worker to adopt new policy weights as soon as its current-version workload drains, without waiting for the rest of the pool.
Migration moves an unfinished trajectory to another worker.

These mechanisms determine when unfinished work runs, where it can resume, and when a worker may change policy versions.
They do not determine which workload shapes should be directed to which heterogeneous accelerators.
That decision depends on how accelerator performance changes with the workload shape that emerges during execution.
 
\section{Scheduling Algorithms}
\label{app:algorithms}

Algorithms~\ref{alg:pacing} and~\ref{alg:tail} formalize the
scheduling mechanisms of \S\ref{sec:design:pacing} and
\S\ref{sec:design:tail}, respectively.
Both maintain two shared structures:
a pending set~$\mathcal{P}_v$ of trajectories awaiting placement,
indexed by policy version~$v$,
and the resident set~$\mathcal{A}_w$ of trajectories currently
hosted on each worker~$w$.
We write $|r|$ for the context length of trajectory~$r$
and $H_w$ for the KV-cache headroom of worker~$w$.

\heading{Pacing (Algorithm~\ref{alg:pacing}).}
The pacing algorithm additionally exploits prompt-level
KV-prefix sharing.
Each trajectory carries a prompt identifier~$\pi(r)$;
a sticky table~$S\colon \pi \to \mathcal{W}$ records the worker
that most recently served each prompt.
A worker is \emph{feasible} for~$r$ if, after placement, its
headroom still exceeds the floor plus a per-trajectory decode reserve:
\[
  \mathcal{F}(r) =
  \left\{w \in \mathcal{W} :
    H_w^{\mathrm{after}}(r) \geq
    H_{\min} + (|\mathcal{A}_w|{+}1)\,\delta\right\}.
\]

\begin{algorithm}[t]
\caption{Pacing: Headroom Restoration and Prefix-Affine Ascending
         Placement (\S\ref{sec:design:pacing})}
\label{alg:pacing}
\begin{algorithmic}[1]
\small
\Require Worker pool $\mathcal{W}$; pending pool $\mathcal{P}_v$;
         headroom floor $H_{\min}$; decode reserve $\delta$;
         sticky table $S$
\algphase{Phase 1: Restore headroom by evicting long residents.}
\ForAll{$w \in \mathcal{W}$ \textbf{such that} $H_w < H_{\min}$}
  \While{$H_w < H_{\min}$ \textbf{and} $\mathcal{A}_w \neq \emptyset$}
    \State $r \leftarrow \arg\max_{r \in \mathcal{A}_w}|r|$
           \Comment{longest first}
    \State evict $r$ from $w$ and add $r$ to $\mathcal{P}_v$
  \EndWhile
\EndFor
\algphase{Phase 2: Place pending trajectories from shortest to longest.}
\State sort $\mathcal{P}_v$ in ascending order of $|r|$
       \Comment{shortest first}
\ForAll{$r \in \mathcal{P}_v$ in sorted order}
  \State compute $\mathcal{F}(r)$
  \If{$\mathcal{F}(r) \neq \emptyset$}
    \If{$S[\pi(r)]$ is defined \textbf{and}
        $S[\pi(r)] \in \mathcal{F}(r)$}
      \State $w^* \leftarrow S[\pi(r)]$
             \Comment{prefix-affine}
    \Else
      \State $w^* \leftarrow
             \arg\min_{w \in \mathcal{F}(r)}
             H_w^{\mathrm{after}}(r)$
             \Comment{tight fit}
    \EndIf
    \State place $r$ on $w^*$ and remove $r$ from $\mathcal{P}_v$
    \State $S[\pi(r)] \leftarrow w^*$
  \EndIf
\EndFor
\end{algorithmic}
\end{algorithm}

\heading{Concentration (Algorithm~\ref{alg:tail}).}
The concentration algorithm identifies the lightest-loaded worker as
its evacuation source via a lexicographic key over affinity rank,
resident count, and total context load:
\[
  \kappa(w) =
  \bigl(\rho(w),\;|\mathcal{A}_w|,\;
        \sum_{r \in \mathcal{A}_w}|r|\bigr).
\]

\begin{algorithm}[t]
\caption{Residual-Tail Concentration (\S\ref{sec:design:tail})}
\label{alg:tail}
\begin{algorithmic}[1]
\small
\Require Worker pool $\mathcal{W}$; pending set $\mathcal{P}_v$;
policy version $v$; headroom floor $H_{\min}$; affinity rank $\rho$
\Statex
\State $\mathcal{W}_v \leftarrow
       \{w \in \mathcal{W}: w \text{ serves version } v\}$
\State $\mathit{source} \leftarrow
       \arg\min_{w \in \mathcal{W}_v} \kappa(w)$
\While{$\mathcal{A}_{\mathit{source}} \neq \emptyset$}
    \If{$\mathcal{P}_v = \emptyset$}
        \State $r \leftarrow
               \arg\max_{r \in \mathcal{A}_{\mathit{source}}}|r|$
        \State evict $r$ from $\mathit{source}$ and add $r$ to
               $\mathcal{P}_v$
               \Comment{longest first}
    \EndIf
    \State sort $\mathcal{P}_v$ in descending order of $|r|$
           \Comment{longest first}
    \State $\mathcal{P}_v^{\mathrm{next}} \leftarrow \emptyset$
    \State $\mathit{progress} \leftarrow \mathsf{false}$
    \State order $\mathcal{W}_v \setminus \{\mathit{source}\}$ by
           $(\rho(w)\ \mathsf{desc}, H_w\ \mathsf{desc})$
    \ForAll{$r \in \mathcal{P}_v$ in sorted order}
        \State $\mathit{placed} \leftarrow \mathsf{false}$
        \ForAll{$w \in \mathcal{W}_v \setminus \{\mathit{source}\}$
                in the above order}
            \While{$H_w^{\mathrm{after}}(r) < H_{\min}$
                   \textbf{and} $\mathcal{A}_w \neq \emptyset$}
                \State $s \leftarrow
                       \arg\min_{s \in \mathcal{A}_w}|s|$
                \State evict $s$ from $w$ and add $s$ to
                       $\mathcal{P}_v^{\mathrm{next}}$
                       \Comment{shortest first}
            \EndWhile
            \If{$H_w^{\mathrm{after}}(r) \geq H_{\min}$}
                \State place $r$ on $w$
                \State $\mathit{placed} \leftarrow \mathsf{true}$
                \State $\mathit{progress} \leftarrow \mathsf{true}$
                \State \textbf{break}
            \EndIf
        \EndFor
        \If{$\neg\mathit{placed}$}
            \State add $r$ to $\mathcal{P}_v^{\mathrm{next}}$
        \EndIf
    \EndFor
    \State $\mathcal{P}_v \leftarrow \mathcal{P}_v^{\mathrm{next}}$
           \Comment{next round}
    \If{$\neg\mathit{progress}$}
        \State wait for a completion or capacity-changing event
    \EndIf
\EndWhile
\State advance $\mathit{source}$ to the next policy version
\end{algorithmic}
\end{algorithm}

The ``next policy version'' in line~35
is not necessarily the latest version
in the system.
If an intermediate version
still has residual tail
on workers with weaker affinity
than the advancing worker,
the advancing worker joins that version
and absorbs the tail first,
advancing further
only after no such residual work remains.
This directs available capacity
to the version on the current critical path.

\heading{Affinity rank rationale.}
Algorithm~\ref{alg:tail} uses the affinity rank $\rho(w)$
in two complementary ways.
The worker with the highest~$\rho$
is the preferred placement target
for residual-tail trajectories,
concentrating long-context work
onto hardware suited for it.
The worker with the lowest~$\rho$
is selected as the evacuation source,
drains first,
and advances to the next policy version,
becoming the first to receive
large-batch fresh work.
A single ordinal ranking
therefore simultaneously matches
the long-context tail
to the hardware best equipped to serve it
and routes fresh work
to the hardware that benefits most
from large active batches.
 
The rank is ordinal rather than cardinal.
Estimating per-move benefit
would require predicting
each trajectory's remaining demand,
which is the circular dependency
identified in \S\ref{sec:principles:circular}.
An ordinal rank sidesteps this
by exploiting a property
that does not require prediction:
each hardware type's relative strength
across different workload shapes
is stable during execution.
A fixed ranking based on
these measured strengths
therefore remains valid
without per-move estimation.
This parallels the pacing mechanism,
which exploits the decode-throughput landscape's
monotonicity to guarantee positive gain
from the \emph{long-out, short-in} transformation.
Both mechanisms use structural properties
to avoid the estimation loop
that \S\ref{sec:principles:circular}
shows to be logically unresolvable.
 
The affinity rank reflects
the comparative advantages of each hardware type
at the two workload shapes
that concentration connects:
the long-context residual tail
and the large-batch fresh work
that follows version advancement.
As Figure~\hyperref[fig:background:hardware-response]{%
  \ref*{fig:background:hardware-response}(a)} shows,
H800 has the most pronounced advantage
at long context lengths,
making it the most effective tail absorber
and earning the highest rank.
A910X has the most pronounced advantage
under large active batches,
so placing it at the lowest rank
ensures it is evacuated first,
advances earliest,
and receives the fresh work
that best matches its strength.
BI-V150's advantage at either end
is less pronounced than H800's or A910X's,
so it occupies the middle rank:
$\rho(\text{H800}) > \rho(\text{BI-V150}) > \rho(\text{A910X})$.
Figure~\ref{fig:evaluation:overtime}
confirms the resulting specialization at runtime,
with H800 workers sustaining
substantially higher mean context lengths
while A910X workers
are the first to advance
and resume large-batch operation.

\heading{Pending set as pace governor.}
During concentration,
workers are evacuated one at a time
and advance to the next policy version
(\S\ref{sec:design:tail},
\emph{Advancing across policy versions}).
If a source drained faster
than the remaining workers
could absorb displaced trajectories,
the tail workers
would bear a disproportionate share
of the residual workload.
Their decode throughput would degrade,
prolonging the very tail
that concentration aims to drain.
Algorithm~\ref{alg:tail} prevents this
through a single guard:
a trajectory is evicted from the source
only when the pending set is empty (line~4).
Each eviction may trigger
a placement cascade.
The evicted trajectory lands
on a high-affinity target,
which may evict its shortest residents
to make room (lines~14--16),
and those displaced residents
enter the pending set
for a subsequent placement pass.
The next eviction from the source
waits until the entire cascade resolves.
Evacuation therefore proceeds
one cascade at a time,
tying the source's drain rate
to the pool's demonstrated absorption capacity.
 
This self-regulation avoids
predicting absorption capacity.
In multi-turn reinforcement learning,
a trajectory's remaining lifetime
depends on how many tool calls remain
and how long each environment interaction takes.
Neither quantity is observable
at scheduling time,
and estimating when a target
will complete its current residents
and free headroom for new work
would reintroduce
the circular dependency
of \S\ref{sec:principles:circular}.
The pending set sidesteps this
by using a single observable condition
$\mathcal{P}_v = \emptyset$
as the sole evacuation gate,
replacing completion-time prediction
with empirical confirmation
that downstream has absorbed
the previous cascade.

\heading{Structural guarantees across mechanisms.}
The pacing transformation,
the affinity rank,
and the pending-set gate
share a common design strategy:
each exploits a structural property
of an observable quantity
to make decisions
whose direction is guaranteed correct
without per-move estimation.
Pacing exploits
the decode-throughput landscape's monotonicity
to guarantee positive gain
from the \emph{long-out, short-in} transformation
without predicting per-move benefit
(\S\ref{sec:principles:gain}).
The affinity rank exploits
hardware comparative-advantage stability
to direct placement
without predicting per-move benefit.
The pending-set gate exploits
the observability of $\mathcal{P}_v = \emptyset$
to pace evacuation
without predicting absorption rate.
In each case,
the structural guarantee
lets the scheduler act
without entering the estimation loop
that \S\ref{sec:principles:circular}
shows to be logically unresolvable.
 
Together the mechanisms form
a continuous cycle across policy versions.
Pacing reshapes the workload
while fresh work is available;
concentration consolidates the tail
and paces version advancement
when fresh work is exhausted;
each advancing worker re-enters pacing
for the new version,
restoring the conditions
for the \emph{long-out, short-in} transformation.
The pending set connects both phases
through a shared interface of 
departure, placement, and headroom.
It acts as a transit station during pacing
and as a pace governor during concentration.
\section{Experimental Configuration}
\label{app:eval-config}
 
Table~\ref{tab:eval-config-homo} lists the device allocations
for homogeneous experiments.
The staleness parameter $\eta$
and decode reserve $\delta$
vary by model size,
as listed in the last two columns.
All configurations ensure
that the trainer has spare capacity
and rollout is the throughput bottleneck.
For Qwen3-8B and Qwen3-14B,
this is achieved by provisioning the trainer
with a larger device share.
Qwen3-32B requires TP${\times}$PP${=}8{\times}2{=}16$
trainer devices at minimum,
forcing a 1{:}1 rollout-to-trainer device ratio
at both pool sizes.
At this ratio,
the device split alone
does not ensure that 
rollout is the actual throughput bottleneck.
Instead, a higher $\eta$
increases the number of concurrent policy versions,
providing enough fresh work
to keep the rollout pool fully utilized
and making rollout the sustained bottleneck.

\begin{table}[tbp]
  \centering
  \setlength{\abovecaptionskip}{4pt}
  \setlength{\belowcaptionskip}{5pt}
  \caption{Device allocation and scheduling parameters
  for homogeneous (H800) experiments.}
  \label{tab:eval-config-homo}
  \normalsize
  \setlength{\tabcolsep}{3.2pt}
  \renewcommand{\arraystretch}{1.16}
  \begin{tabular*}{\columnwidth}{l@{\extracolsep{\fill}}r c c c c r r}
    \toprule
    & & 
    \multicolumn{2}{c}{Rollout}
    & \multicolumn{2}{c}{Trainer}
    & & \\
    \cmidrule(lr){3-4} \cmidrule(lr){5-6}
    Model & Total
          & Dev. & TP
          & Dev. & TP${\times}$PP
          & $\eta$ & $\delta$ \\
    \midrule
    Qwen3-8B  & 16 & 8  & 1 & 8  & 8${\times}$1 & 1 & 512 \\
    Qwen3-8B  & 32 & 16 & 1 & 16 & 8${\times}$1 & 2 & 512 \\
    Qwen3-14B & 24 & 8  & 2 & 16 & 8${\times}$1 & 1 & 1{,}024 \\
    Qwen3-14B & 48 & 16 & 2 & 32 & 8${\times}$1 & 2 & 1{,}024 \\
    Qwen3-32B & 32 & 16 & 4 & 16 & 8${\times}$2 & 3 & 2{,}048 \\
    Qwen3-32B & 64 & 32 & 4 & 32 & 8${\times}$2 & 4 & 2{,}048 \\
    \bottomrule
  \end{tabular*}
\end{table}

\section{Baseline Descriptions}
\label{app:baselines}

We compare \sysname{} against five strategies
that span the design space of rollout coordination
(\S\ref{sec:evaluation:setup}).
Because the implementation exposes rollout coordination
as a plug-and-play strategy,
\sysname{} and every baseline reuse
the same underlying rollout and training framework.
The comparison therefore varies
the coordination strategy
while keeping the framework fixed.
 
\begin{itemize}
    \item \textbf{Sync.}
    We use a synchronous reference in which rollout and training execute sequentially.
    After each training stage, the updated policy weights are synchronized to the rollout workers before the next rollout stage, so trajectories are generated and consumed without policy-version staleness.
 
    \item \textbf{One-off.}
    One-step off-policy execution overlaps rollout with training using trajectories from the previous policy version, bounding staleness to one policy step~\cite{zhong2025streamrl}.
 
    \item \textbf{DORA$^{*}$.}
    We implement the workload-aware resource-allocation component of DORA, which reallocates rollout workers across policy versions according to their current workloads~\cite{hu2026dora}.
    In our implementation, this reallocation is performed at policy-weight synchronization points.
    The asterisk indicates that this is a partial implementation and does not represent the full DORA system.
 
    \item \textbf{Laminar.}
    Laminar introduces a trajectory-level asynchronous rollout policy, where workers progress independently and obtain newer policy weights after completing their current generation batches~\cite{sheng2026laminar}.
    Its repack policy consolidates unfinished trajectories from underutilized workers onto fewer workers within the same policy-version group.
 
    \item \textbf{AReaL.}
    We implement AReaL's partial-rollout policy, which interrupts ongoing trajectory generation when the trainer publishes new policy weights and resumes unfinished trajectories with the updated weights~\cite{fu2026areal}.
    The interrupted prefix is recomputed under the new weights, so a trajectory may contain segments generated by multiple policy versions.
\end{itemize}
 
\section{Heterogeneous Pool Analysis}
\label{app:hetero-analysis}

This appendix expands the heterogeneous evaluation
in \S\ref{sec:evaluation:end-to-end}
with mechanistic analysis
that connects the observed results
to the hardware throughput profiles
characterized in \S\ref{sec:background:hardware}
and the scheduling mechanisms
described in \S\ref{sec:design:pacing}
and \S\ref{sec:design:tail}.

\begin{figure}[t]
  \centering
  \includegraphics[width=\columnwidth]{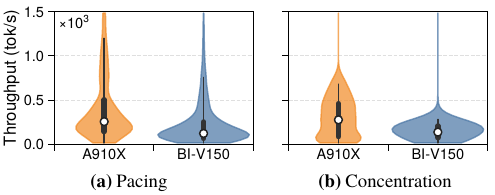}
  \caption{Per-accelerator runtime throughput distributions
  during the pacing and concentration phases
  on homogeneous pools (32A, 32B).
  During pacing, reshaping keeps both accelerator types
  in a favorable operating range,
  while A910X sustains higher throughput.
  During concentration,
  both accelerator types serve the residual long-context workload
  and the throughput gap persists.}
  \label{fig:app:hetero:phase-throughput}
\end{figure}

\heading{Composition sensitivity and phase throughput.}
Relative to the H800,
whose throughput is relatively robust
to workload-composition changes,
A910X and BI-V150 exhibit two distinct
and representative sensitivities.
A910X is more sensitive to active batch size,
whereas BI-V150 is more sensitive
to within-batch context-length skew.
These contrasting responses
are visible in the workload-dependent profiles
of Figure~\ref{fig:background:hardware-response}.
The 32A and 32B configurations
in Figure~\ref{fig:evaluation:hetero}
show \sysname{} improving
both throughput and latency over the baselines,
eliminating the throughput--latency trade-off
visible in the H800 comparison.
The phase-level distributions in
Figure~\ref{fig:app:hetero:phase-throughput}
explain both the improvement
and the remaining difference between the two pools.

Figure~\hyperref[fig:app:hetero:phase-throughput]{%
  \ref*{fig:app:hetero:phase-throughput}(a)%
}
shows per-accelerator throughput during pacing.
Compared with the runtime distributions
under one-time placement
in Figure~\hyperref[fig:background:hardware-response]{%
  \ref*{fig:background:hardware-response}(c)%
},
both accelerator types operate
in a higher and more concentrated throughput range.
The \emph{long-out, short-in} transformation
therefore addresses the two sensitivities
in their respective favorable directions.
On A910X, it sustains the active batch size;
on BI-V150, it reduces within-batch context-length skew.
The pacing throughput of A910X exceeds
that of BI-V150 because,
at the same TP${=}2$,
each A910X worker provides twice
the aggregate HBM capacity
(128\,GB versus 64\,GB).
This larger capacity lets A910X workers sustain larger active batches
and achieve higher per-worker throughput,
yielding the higher aggregate decode throughput
of the 32A pool in Figure~\ref{fig:evaluation:hetero}.

This difference determines how the two pools
move through a complete pacing--concentration cycle.
Under the same workload distribution,
fresh-work supply,
and admission budget,
the higher pacing throughput of 32A
consumes the available fresh work
in less wall-clock time
and reaches the concentration phase sooner.
With the same residual workload,
the earlier transition lets concentration
assign more decode resource
to each remaining trajectory.
Figure~\hyperref[fig:app:hetero:phase-throughput]{%
  \ref*{fig:app:hetero:phase-throughput}(b)%
}
shows the corresponding distributions during concentration.
Both pools are serving the residual long-context workload,
and the A910X distribution remains centered
at higher throughput than the BI-V150 distribution.
The phase-level advantage of 32A
therefore continues while the tail is drained.

\heading{Aggregate throughput and non-resident time.}
Table~\ref{tab:app:nonresident-fraction}
reports the non-resident fraction,
the share of a long-context trajectory's lifecycle
spent in the pending set
rather than being resident and actively decoding.
It is the trajectory-level view
of the yield-and-return lifecycle
examined in \S\ref{sec:evaluation:end-to-end}.
A trajectory is resident while it decodes,
enters the pending set when it is yielded,
and becomes resident again
when a worker with sufficient headroom
can resume it.

Across pool configurations,
those with higher aggregate decode throughput
in Figure~\ref{fig:evaluation:hetero}
exhibit lower non-resident fractions
at the reported percentiles.
The 32A--32B comparison is the clearest example.
The non-resident fraction is lower on 32A
at every percentile,
with the gap widening from P50
to the upper tail.
Thus the aggregate-throughput difference
is also a difference in how long
long-context trajectories remain without active decode.

The mechanism follows from the resident and pending roles.
A displaced trajectory remains in the pending set
until some worker frees sufficient KV headroom.
Higher aggregate decode throughput
lets workers process their resident trajectories faster,
expose headroom sooner,
and return pending trajectories
to the resident state sooner.
Together with the earlier transition
and the larger decode allocation
per trajectory during concentration,
this reduces non-resident time
and the resulting long-tail latency.
The table therefore links
the phase-throughput difference
to trajectory-level behavior.

\begin{table}[t]
  \centering
  \setlength{\abovecaptionskip}{4pt}
  \setlength{\belowcaptionskip}{5pt}
  \caption{Non-resident fraction of long-context trajectories
  across pool configurations under \sysname{}.}
  \label{tab:app:nonresident-fraction}

  \normalsize
  \setlength{\tabcolsep}{3.2pt}
  \renewcommand{\arraystretch}{1.16}

  \begin{tabular*}{\columnwidth}{l@{\extracolsep{\fill}}c c c c}
    \toprule
    Pool & P25 & P50 & P75 & P95 \\
    \midrule
    32A      & 0.21\% & 0.89\% &  5.42\% & 39.91\% \\
    32B      & 0.29\% & 1.61\% & 37.14\% & 80.94\% \\
    16A16B   & 0.18\% & 0.79\% &  7.05\% & 68.05\% \\
    16A8H    & 0.19\% & 0.88\% &  7.27\% & 50.72\% \\
    16B8H    & 0.34\% & 1.58\% & 17.76\% & 60.55\% \\
    16A16B8H & 0.17\% & 0.64\% &  2.80\% & 40.31\% \\
    \bottomrule
  \end{tabular*}
\end{table}

\end{document}